%% file: main.tex
\pdfoutput=1

\documentclass[10pt,journal,compsoc]{IEEEtran}
\usepackage{listings}
\usepackage{tcolorbox}
\usepackage[caption=false,font=normalsize,labelfont=sf,textfont=sf]{subfig}
\usepackage{cite}
\definecolor{lightred}{rgb}{1.0, 0.8, 0.8}
\definecolor{lightgreen}{rgb}{0.8, 1.0, 0.8}

\input{def}
\begin{document}

\title{Identifying and Mitigating API Misuse in Large Language Models}

\author{Terry Yue Zhuo,
        Junda He,
        Jiamou Sun,
        Zhenchang Xing,
        David Lo,
        John Grundy,
        and~Xiaoning Du% <-this % stops a space
\thanks{Corresponding author: Xiaoning Du}
\thanks{T.Y.Zhuo and J.Grundy and X.Du are with Monash University, Australia.
E-mail: terry.zhuo@monash.edu, john.grundy@monash.edu, xiaoning.du@monash.edu}% <-this % stops a space
\thanks{T.Y.Zhuo, J.Sun, and Z.Xing are with CSIRO's Data61, Australia.
E-mail: frank.sun@data61.csiro.au, zhenchang.xing@data61.csiro.au}% <-this % stops a space
\thanks{J.He and D.Lo are with Singapore Management University, Singapore.
E-mail: jundahe@smu.edu.sg, davidlo@smu.edu.sg}% <-this % stops a space
\thanks{Z.Xing is also with Australian National University, Australia.}
}

\markboth{IEEE Transactions on Software Engineering}
{Shell \MakeLowercase{\textit{et al.}}: Bare Demo of IEEEtran.cls for Computer Society Journals}

\IEEEtitleabstractindextext{
\begin{abstract}
\justifying
API misuse in code generated by large language models (LLMs) presents a serious and growing challenge in software development. While LLMs demonstrate impressive code generation capabilities, their interactions with complex library APIs are often error-prone, potentially leading to software failures and vulnerabilities.
In this paper, we conduct a large-scale study of API misuse patterns in LLM-generated code, analyzing both method selection and parameter usage across Python and Java, using three representative LLMs (StarCoder-7B, Qwen2.5-Coder-7B, and GitHub Copilot). Based on extensive manual annotation of 3,209 method-level and 3,492 parameter-level misuses, we identify and categorize four recurring misuse types by building on and refining prior API misuse taxonomies.
Our evaluation of three widely used LLMs, StarCoder-7B, Qwen2.5-Coder-7B, and GitHub Copilot, reveals persistent challenges in API usage, particularly hallucination and intent misalignment.
To address these issues, we propose Dr.Fix, an LLM-based automatic repair approach guided by our taxonomy. Dr.Fix improves repair accuracy compared to baseline prompting and existing repair methods, with gains of up to 38.4 BLEU and up to 72 more exact matches on benchmark datasets.
This work offers important insights into the current limitations of LLMs in API usage and provides insights into current limitations and points to directions for improving automated misuse repair in code generation systems.
\justifying
\end{abstract}

\begin{IEEEkeywords}
Code Generation, Large Language Model, API Misuse, Automatic Program Repair, Empirical Software Engineering
\end{IEEEkeywords}}

\maketitle

\section{Introduction}

Recent advances in large language models (LLMs) have significantly enhanced the sophistication of automatic code generation, as these models are trained on vast datasets of source code~\cite{touvron2023llama, achiam2023gpt, li2023starcoder, lozhkov2024starcoder, muennighoff2024octopack, zhauo-etal-2025-nlp}. However, according to previous work, LLMs may still generate suboptimal or incorrect API usage in code generation despite being exposed to extensive code during training~\cite{zhuo2024bigcodebench}. Library APIs have been pivotal in software development, enabling complex functionalities across various domains, including network communication, data visualization, and scientific computation~\cite{zhong2017empirical}. Moreover, as libraries evolve by introducing new features, deprecating old ones, or changing usage patterns, LLMs may struggle to adapt to these changes~\cite{islah2024gitchameleon, kuhar2024libevolutioneval}. With over 590,000 external libraries\footnote{\url{https://pypi.org/}}
 in languages like Python, using the right API correctly can be a considerable challenge, even for experienced developers.

Despite these advancements, existing studies on API misuse in LLM-generated code have focused primarily on controlled, synthetic scenarios. For example, \cite{zhong2024can} investigated the quality of code generated by several LLMs using StackOverflow-style programming questions, finding that GPT-4 generated code with API misuses in 62\% of cases when tested on 18 widely used Java APIs. Similarly, \cite{mousavi2024investigation} studied API misuse in LLMs by examining security-related programming questions in Java, revealing that LLMs misused APIs in approximately half of the tasks. A more recent error analysis of 1,000 samples from existing benchmarks identified three primary error types, ``AttributeError'', ``TypeError'', and ``ValueError'', commonly associated with API misuse in LLM-generated code. While these studies have demonstrated that specific LLMs can generate code with API misuse on limited libraries, APIs, and programming languages, their findings may not generalize to real-world programming contexts due to their reliance on human-curated questions and scenarios.

A more common scenario in real-world software development involves LLMs integrated within IDE environments, which assist developers in continuing or completing existing code~\cite{xu2022ide}. LLMs are often prompted with an incomplete code snippet and asked to suggest an appropriate API call. In this context, LLMs face distinct challenges: They may erroneously hallucinate functions from other libraries, suggest irrelevant or incorrect APIs, or specify invalid parameters, even when familiar with the target library. These context-dependent completion tasks present challenges different from those of generating standalone code snippets in response to specific queries. While previous research has studied API misuse of LLMs on task-oriented synthetic datasets~\cite{zhong2024can,mousavi2024investigation}, there remains a critical gap in understanding whether LLMs exhibit similar patterns of API misuse in real-world code completion. Specifically, it remains unclear whether LLMs make the same types of mistakes in API usage as human developers do. Moreover, investigating API misuse patterns across both open-source and closed-source models is essential, as open-source models often lag behind their closed-source counterparts in terms of performance. Current approaches to mitigating API misuse primarily focus on automatic detection using predefined misuse types and API documentation. However, these methods face significant limitations when applied to LLM-based code generation, which involves millions of APIs and varying contexts. Even when misuse is detected correctly, rule- and program-analysis-based automatic program repair (APR) techniques often fail to fix misused APIs due to their limited semantic understanding~\cite{kechagia2021evaluating,zhang2023evaluating}. This suggests that LLM-based APR approaches, which can leverage the semantic understanding of code, may be more suitable for addressing such issues.

In this work, we investigate how LLMs misuse APIs and examine their behavior in real-world code-generation scenarios. Specifically, we evaluate three representative decoder-only LLMs used in IDE settings: StarCoder-7B~\cite{li2023starcoder} and Qwen2.5-Coder-7B, two open-source models for code completion, and Copilot, a closed-source model powered by GPT-4~\cite{achiam2023gpt}.  Both models are widely used in IDE environments such as VSCode\footnote{\url{https://code.visualstudio.com/}}
. While previous human-centric API misuse research often studies patches from fixed commits~\cite{kechagia2021evaluating,zhang2023evaluating}, it is not feasible to treat LLM-generated code in the same way, as LLMs generate outputs in various formats. We sample code snippets from GitHub in Python and Java to systematically assess how these models misuse APIs. We conduct program analysis to localize the API positions and provide the code context before these APIs as input to the models. Unlike existing studies, we assess the correctness of API invocation in two key areas: method selection and parameter usage. The first aspect evaluates whether LLMs can suggest an appropriate API name based on the code context, while the second examines whether LLMs can predict reasonable API parameters. These aspects correspond to typical use cases in IDE environments' code filling~\cite{friedincoder} and left-to-right code completion~\cite{radford2018improving}.

Building on existing taxonomies of API misuse~\cite{wei2024demystifying, he2023python}, we adapt and refine them to the context of LLM-generated code. Specifically, we retain violation types that can be reliably identified from snippet-level context, such as Missing and Redundant, and introduce two additional categories that capture LLM-specific failure modes, namely Intent misuse and Hallucination misuse, while de-emphasizing categories that depend on broader project or version information, such as Replacement and Outdated. Through extensive manual annotation of 3,209 method-level and 3,492 parameter-level cases, we identify four recurring misuse patterns in LLM-generated API code: (1) Intent misuse, where a model selects an API that is syntactically valid but semantically inappropriate for the intended task; (2) Hallucination misuse, where the model produces non-existent or fabricated API methods or parameters; (3) Missing item misuse, where required API methods or parameters are omitted; and (4) Redundancy misuse, where unnecessary API calls or arguments are introduced, often leading to inefficiency or potential errors.

Our systematic study reveals that LLMs struggle significantly with API misuse, particularly hallucination and intent misalignment, with distinct patterns observed between Python and Java. While Copilot and Qwen2.5-Coder generally outperform StarCoder, their shared error patterns indicate common challenges in API usage across models. To address these issues, we propose \underline{D}etect-\underline{r}eason-\underline{Fix} (\texttt{Dr.Fix}), a taxonomy-guided LLM-based APR method. Compared with baseline prompting and prior repair methods, \texttt{Dr.Fix} achieves substantial improvements, with BLEU scores increasing by up to 38.4 points and up to 72 more exact matches across different models and languages.

The key contributions of this research include:
\begin{itemize}
\item An empirical study of API misuse in LLM-generated code completion, conducted on Python and Java using three representative decoder-only LLMs (StarCoder-7B, Qwen2.5-Coder-7B, and Copilot).
\item An adaptation of existing API misuse taxonomies to the LLM setting, refined through the manual annotation of 6,701 misuse cases across method and parameter levels.
\item A publicly available dataset and replication package to support transparency and reproducibility.
\item A taxonomy-guided LLM-based APR approach, \texttt{Dr.Fix}, evaluated against baseline prompting and prior work, showing significant improvements in repair accuracy.
\end{itemize}

The rest of the paper is organized as follows. We outline the background in \autoref{sec:background}. We elaborate on the details of the experimental setting for the case study and introduce the misuse taxonomy in \autoref{sec:case_study}. Based on the annotations on generating three LLMs on Python and Java, we conduct quantitative analysis and share several insights. Then, we propose the \texttt{Dr.Fix} for in-the-wild API misuse repair and evaluate its performance against LLM-based APR on our benchmark in \autoref{sec:mitigation}. To facilitate the future research of API misuse, we provide a replication package~\cite{llm_api_misuse_replication} and plan to release the code and dataset publicly on GitHub upon acceptance.
% However, like humans, LLMs have to learn large-scale source code data to write high-quality programs, and  

\begin{figure}
\begin{lstlisting}[language=Python]
import torch
import torch.nn as nn

class GRUCell(nn.Module):
    """
    A simple implementation of a GRUCell in PyTorch.
    """
    def __init__(self, input_size, hidden_size, bias=True):
        super(GRUCell, self).__init__()
        self.input_size = input_size
        self.hidden_size = hidden_size
        self.bias = bias
        # LLM is expected to complete the next line with an API call, such as nn.Linear
        self.ih = 
\end{lstlisting}
\caption{An example of LLM code completion suggesting a PyTorch API call for the last line.}

\label{fig:teaser}
\end{figure}

\section{Background}
\label{sec:background}

\subsection{Motivation}

Consider the example of LLM code generation shown in \autoref{fig:teaser}. When prompted with an incomplete code snippet, an LLM may suggest an inappropriate API call due to misunderstanding the usage context. Typical errors include: (1) incomplete method calls or those with erroneous parameters, (2) calls to similar but unrelated APIs, (3) extraneous method calls not needed in the current context, (4) incorrect sequencing of method calls for library setup, and (5) incorrect integration of APIs from multiple libraries.

With the rise of support for both code completion and code infilling in modern IDEs such as Cursor\footnote{\url{https://docs.cursor.com/en/tab/overview}}
, it is important to examine how reliably LLMs can generate correct API usage under these two scenarios. Since API calls are the central building blocks of modern software development, even small deviations from correct usage can propagate into functional errors, security issues, or inefficiencies. Therefore, before presenting our evaluation setup, we first scope what we consider to be API misuse in the context of LLM-generated code and how it differs from general programming errors.

\subsection{API Misuse}

APIs are essential components of modern software systems, enabling developers to interact with third-party libraries and services. Correct usage of APIs ensures that functionality is accessed as intended, while misuse can result in unexpected behavior, degraded performance, or even security vulnerabilities~\cite{amann2018systematic}. API misuse typically occurs when developers violate usage constraints, such as required input types, parameter ordering, or call sequences. Unlike general syntax or language-level type errors (e.g., null dereferencing~\cite{bush2000static}), API misuses may not always lead to immediate failure but can still cause subtle, incorrect behavior~\cite{khatchadourian2020empirical}. The causes of API misuse are multifaceted, including incomplete documentation, lack of domain knowledge, and evolving API designs~\cite{lamothe2021systematic}. Consequences range from performance degradation to system failure, particularly in high-stakes domains like healthcare or finance~\cite{mousavi2024investigation}. Prior studies have extensively investigated human-written API misuses through bug-fixing commits and empirical analysis~\cite{schlichtig2022fum, ren2020demystify, galappaththi2024empirical, wei2024demystifying}.

In this work, we adopt the usage constraint framework from Schlichtig et al.~\cite{schlichtig2022fum} and apply the taxonomy proposed by Wei et al.~\cite{wei2024demystifying} and He et al.~\cite{he2023python}, which classifies API misuses based on violation types (e.g., Missing, Redundant) and API elements (e.g., Method, Parameter, Condition). Although originally developed for deep learning libraries, we generalize this taxonomy to broader third-party APIs. We follow a closed coding approach to categorize LLM-generated misuses under this schema. Following \cite{amann2016mubench}, we define an API misuse as an incorrect use of an API that violates its documented contract or commonly expected usage constraints at the level of a specific API element. 
For instance, passing a string where a list is required is classified as an API misuse, even if the code is syntactically valid in Python, since it semantically violates the API’s intended usage. In contrast, general language-level errors such as undefined variables, misspellings, or unrelated type mismatches are excluded unless they directly affect API invocation~\cite{amann2016mubench, wei2024demystifying, he2023python}.
We adapt this taxonomy for the LLM setting and introduce two descriptive refinements, \textit{intent misuse} and \textit{hallucination}, as surface-level subtypes of existing misuse categories. Intent misuse reflects cases where the model selects a valid API element that is semantically incorrect given the task (e.g., using \texttt{vs.abs} instead of a vector magnitude function). Hallucination captures cases where the model generates nonexistent methods or parameters. While both map to known categories like \textit{Replacement} or \textit{Missing–Method}, they are particularly prevalent in LLM outputs and not emphasized in prior human-centric taxonomies. We include them to better reflect distinctive patterns in LLM-generated code without altering the underlying taxonomy.

\subsection{Mitigation of API Misuse}

API usage can be categorized into two primary types: directly calling API methods or instantiating objects from API classes~\cite{nguyen2010graph}. API misuses are defined as violations of the implicit or explicit usage constraints of APIs~\cite{amann2018systematic}. Recently, numerous API-misuse detectors have been developed~\cite{kechagia2019effective,kang2021active,li2021arbitrar}. Generally, there are three types of API misuse detection approaches: static detectors, which identify API misuses through static analysis of source or binary code~\cite{monperrus2013detecting}; dynamic detectors, which detect API misuses through dynamic analysis~\cite{pradel2012leveraging}; and hybrid approaches that combine mining techniques with static analysis~\cite{pradel2012statically}. Regardless of the detection technique, existing methods (1) require API specifications, considering the violation of these specifications as API misuse. However, these specifications are often incomplete and difficult to obtain~\cite{li2021arbitrar}; or (2) do not require explicit API specifications, instead relying on the assumption that the majority usage pattern in a large-scale code corpus represents valid usage. An API is considered misused if it deviates from this majority pattern. The limitation is that this assumption may not always hold, especially for rarely used APIs where the majority pattern does not necessarily reflect valid usage~\cite{kang2021active}. 
More recently, there has been promising work exploring the use of LLMs to repair API misuse by fine-tuning paired commit data~\cite{zhang2023evaluating}. Other approaches, such as~\cite{wei2024demystifying, kechagia2021evaluating}, provide domain-specific or rule-based repair mechanisms. In this work, we compare our proposed method not only against prompting baselines, but also contextualize it with these prior repair techniques, highlighting the advantages and limitations of LLM-based approaches for API misuse repair.

\section{Qualitative Studies}
\label{sec:case_study}

This section presents our qualitative analysis of API misuse in LLM-generated code. We do not propose a new taxonomy; rather, we apply a well-established classification framework from Schlichtig et al.~\cite{schlichtig2022fum}, supplemented with categories from Wei et al.~\cite{wei2024demystifying} and He et al.~\cite{he2023python}, to annotate API misuse cases. These taxonomies define violations in terms of location (method, parameter, condition) and type (missing, redundant, replacement, outdated). Our goal is to assess whether LLMs exhibit similar misuse patterns to human developers, and where they differ.

We use this framework to analyze outputs from LLMs in realistic code completion tasks, focusing on API correctness in two common generation scenarios.

\subsection{Research Questions}

We wanted to answer the following key research questions about API misuse by current LLMs:

\textbf{RQ1: How well do LLMs handle API invocation in code generation?} We first investigate how reliably LLMs can generate correct API usages in two modern IDE scenarios: code completion, where models predict the next line or token sequence, and code infilling, where models fill in masked elements such as method names or parameters. These objectives reflect real-world developer interactions, where models are often used to suggest API methods or complete parameter lists within partially written code. Our design builds on traditional studies of code completion~\cite{proksch2016evaluating, raychev2014code} as well as recent work on infilling techniques~\cite{friedincoder, li2023starcoder, roziere2023code}, ensuring that our evaluation setup aligns with both established and emerging IDE practices.

\textbf{RQ2: How do LLMs misuse APIs in the wild?} As automatic evaluation metrics can only reflect the overall correctness, we then wanted to understand the specific misuse patterns of LLMs from the human perspective.

\subsection{Evaluation Elements}

To invoke an API correctly, LLMs are expected to predict both the correct API method name and pass valid arguments for the API parameters, where the objectives are similar to the human-written code~\cite{schlichtig2022fum}. We design the following two scenarios to evaluate the correctness of the generated API elements:

\paragraph{\textbf{API Method Infilling}} In this evaluation setup, we mask out the API method name in a code snippet and ask the models to infer the appropriate method, similar to type inference~\cite{cassano2023type,peng2023generative}. This evaluation assesses whether the models can comprehend the intent of the code context and recommend a suitable method name based on the imported packages and the parameters' context.

\begin{figure}[h]
\begin{lstlisting}[
    language=Python, 
    escapechar=!
]
import requests

def fetch_data(url):
    # model will fill in the mask by predicting an API name
    response = requests.!\textbf{[MASK]}!(url) 
    return response.json()
\end{lstlisting}
\caption{Example for API Method Infilling: The model should predict `get` as the appropriate method name.}
\label{fig:api_method_inference}
\end{figure}

As shown in \autoref{fig:api_method_inference}, the model is expected to predict `get` as the appropriate method name.

\paragraph{\textbf{API Parameter Completion}} In this case, rather than focusing on predicting the method name, we evaluate the models' ability to generate valid and contextually appropriate parameter values for a given API method. Here, the model is provided with a code snippet where the API method name is already specified, and it is tasked with generating the appropriate argument(s) within the method call. All surrounding context after the method name is masked, requiring the model to infer correct usage based on its knowledge of the API. The models are expected to complete the arguments correctly by recalling the parameter names and types associated with the API method and identifying potential variables as valid arguments.

\begin{figure}[h]
\begin{lstlisting}[language=Python, 
    escapechar=!]
import requests

def fetch_data():
    url = "http://example.com/api/data"
    response = requests.get(!\textbf{[MASK]}!)
    return response.json()
\end{lstlisting}
\caption{Example for API Parameter Completion: The model should predict `url' as the appropriate parameter.}
\label{fig:api_parameter_generation}
\end{figure}

As shown in \autoref{fig:api_parameter_generation}, the model is expected to predict `url` as the appropriate parameter.

For evaluation, Exact Match (EM) is computed at the level of the masked code elements (API method name or parameters), rather than the entire snippet, to directly capture API correctness. In addition to BLEU and CodeBERTScore, we also discuss precision and recall at the API-element level to better reflect the accuracy of API invocations.

\subsection{Models}

LLMs differ in architecture, size, and pretraining tasks. In practice, decoder-only models are preferred in IDE environments, where developers rely on them for code completion and comment generation~\cite{svyatkovskiy2020intellicode}. They are also used to summarize code for comprehension or generate test cases for target snippets. Current IDEs such as VSCode and Cursor\footnote{\url{https://www.cursor.com/}}
 primarily deploy decoder-only base models that directly predict the next tokens without instruction tuning. This design supports fast inference and seamless integration into interactive editing workflows~\cite{izadi2024language,takerngsaksiri2024students,semenkin2025full}. Open-source LLMs are typically executed locally to maximize inference speed and customization flexibility (e.g., domain-specific fine-tuning)~\cite{fu2024serverlessllm}, while closed-source LLMs offer higher performance but can only be accessed via hosted APIs.
To capture representative behaviors of LLMs commonly used in IDEs, we began our experiments in early 2024 and selected three widely used models for code generation~\cite{Wiggers2023StarCoderFreeModel,hui2024qwen2,IBM2023watsonxMainframe,Dohmke2023CopilotTransformsPlatform}: StarCoder-7B, an open-source decoder-only model from Hugging Face; Qwen2.5-Coder-7B, an open-source model from Alibaba’s Qwen2.5 family; and GitHub Copilot, a closed-source model powered by OpenAI’s GPT-4. These models collectively represent both open and closed ecosystems, balancing between transparency, accessibility, and real-world deployment in IDE workflows. Although newer versions such as StarCoder2 have since been released, our selected models remain representative of IDE-integrated LLMs at the time of study.
 
\textbf{\underline{StarCoder}} is the first fully permissive LLM family trained on code by Hugging Face, outperforming popular LLMs and matching or surpassing closed-source models like `code-cushman-001` from OpenAI (the original Codex model that powered early versions of GitHub Copilot). With a context length of over 8,000 tokens, the StarCoder models can process more input than any other open-source LLM, enabling a wide range of interesting applications. In addition, the model supports both left-to-right code completion and fill-in-the-middle code prediction. Unlike previous LLMs that are only trained on common languages, StarCoder supports 86 programming languages from high resources (e.g., Java) to low resources (e.g., Dart). Furthermore, StarCoder provides three different model sizes, from 1B to 15B. The 1B model maximizes the inference speed, and the 15B model optimizes the generation quality. In our study, we chose StarCoder-7B, which is the best model balancing between accuracy and efficiency.

\textbf{\underline{Qwen2.5-Coder-7B}} is a recent open-source code generation model from Alibaba’s Qwen2.5 series. Like StarCoder, it adopts a decoder-only transformer architecture optimized for next-token and fill-in-the-middle prediction tasks. It is pretrained on a large corpus of multilingual source code and natural language, supporting over 90 programming languages. Compared with StarCoder, Qwen2.5-Coder incorporates more recent repositories and improved tokenizer coverage for Python and Java, making it well-suited for IDE-style code completion without requiring instruction tuning.

\textbf{\underline{Copilot}} is a closed-source small code suggestion model distilled from OpenAI's GPT-4, initially launched in 2022. While the model is further fine-tuned for coding tasks, it has a much better understanding of natural language context than open-source LLMs like StarCoder. More than 46\% of developers’ code files, on average, are reportedly generated by GitHub Copilot. Similar to StarCoder, Copilot offers two generation paradigms: code completion and code infilling. Besides, Copilot utilizes a client-side model to reduce the frequency of unwanted suggestions when they might prove disruptive to a developer’s workflow, helping it better respond to each developer using it. In this work, we use the Copilot version, which started in 2024, and manually paste code snippets into VSCode for model inference.

\subsection{Hyperparameters}

To align real-world software development in the IDE with LLMs, we use the default hyperparameters specified in each VSCode extension. Regarding StarCoder, we follow the configurations of Hugging Face's \texttt{llm-ls}\footnote{\url{https://github.com/huggingface/llm-ls}}, a Language Server Protocol server leveraging LLMs for code completion. Specifically, we use the temperature of 0.2, the maximum new token number of 150, and the top\_p value of 0.95. For Qwen2.5-Coder-7B, we use the recommended hyperparameters, with the temperature of 0.3 and the top\_p value of 0.95. For GitHub's Copilot model, we use the official v1.252.0 extension in VSCode 1.95.3 and prompt the model with the code snippet in a single file. Specifically, we directly query the models for autocompletion without any customized instructions.

\subsection{Dataset}
\label{subsec:data}

We describe how we construct the dataset for our qualitative study. We leverage The Stack~\cite{kocetkovstack} data, the largest and most up-to-date permissively licensed GitHub corpus containing source code files in 358 programming languages from GitHub repositories. Due to the high data quality resulting from the comprehensive filtering mechanisms, The Stack dataset has been partially utilized by a series of LLMs for pretraining. Specifically, we use v1.2, the latest version that opted out of the requested and malicious data, and exclude the subset used to train StarCoder to avoid data contamination issues~\cite{sainz2023nlp}.

To conduct a thorough investigation, we selected two popular programming languages, Python and Java. As each code file can contain multiple library APIs, we localize all the positions of APIs (methods and parameters) via static analysis. It is important to note that there could be alternative signatures for the same APIs, particularly where methods share the same name but differ in their parameters (e.g., overloading or different argument types). Additionally, we consider APIs that provide the same functionality but reside in different paths or namespaces due to version differences. To account for this, we ensure that all distinct signatures are included in our dataset. For example, an API could appear in various forms depending on the number and types of parameters used. To ensure comprehensive coverage of different programming scenarios, we include only one sample for each fully qualified-named API signature in our dataset, which helps us capture the diversity of API usage across various contexts. Furthermore, there are many duplicated APIs invoked in various code snippets. To ensure our examination covers diverse programming scenarios in the wild, we restrict the dataset to at most one instance of each unique, fully-qualified API signature \textbf{per file}. This prevents highly frequent APIs (e.g., \texttt{numpy.array}, \texttt{requests.get}) from dominating the dataset while still allowing us to capture multiple usages of the same API across different projects and contexts. 
In total, we randomly sample 3,000 unique APIs for both Python and Java, respectively, based on their distinct fully qualified names. The APIs are collected from 631 Python and 486 Java repositories, covering 4,641 Python packages and 6,258 Java packages identified through import statements. This broad collection ensures coverage of diverse libraries and API usage contexts, reflecting a wide range of real-world development practices. For each API usage sample, we treat the code before the API method as a prefix and the parenthesized parameters as a suffix for \textit{API Method Infilling}, and the code before the API parameter as a prefix for \textit{API Parameter Completion}, resulting in 36,000 generated samples.

Although The Stack applies rigorous filtering, we acknowledge that human-written code may still contain errors; therefore, we treat it as an approximate ground truth, assuming that the majority of examples represent correct API usage while recognizing that occasional misuses or stylistic inconsistencies may introduce noise. Since our dataset emphasizes frequently used APIs, the findings may not fully generalize to rarely used or newly introduced APIs, which we discuss further in threats to validity in \autoref{sec:construct_validity}.

\subsection{Evaluation Metrics}
\label{subsec:metric}

As we randomly sample our dataset from a raw GitHub code corpus without environment configurations (e.g., dependencies) and oracles, it is hard to assess the code quality via compilation and test cases. Following prior works~\cite{hu2024active, liu2023codegen4libs, ma2024compositional}, we use BLEU~\cite{papineni2002bleu}, CodeBERTScore~\cite{zhou2023codebertscore}, and Exact Match (EM) as evaluation metrics to determine the (un)likelihood that the models can invoke the same APIs as human developers. Since automatic metrics have known limitations, we further complement them with element-level precision and recall, as well as manual analysis of misuse categories.

\textbf{BLEU} BLEU calculates the percentage of 4-gram overlap between the reference fixed code and the candidate code generated by the models. For simplicity, we refer to it as BLEU in our work. The candidate fixed code represents the code generated by the models, while the reference fixed code represents the ground-truth code written by developers. BLEU is defined as follows:

\[
\text{BLEU} = \text{BP} \cdot \exp\left(\sum_{n=1}^{4} w_n \log p_n\right)
\]

where \(p_n\) is the precision of n-grams, \(w_n\) is the weight for n-grams (usually equal), and \(\text{BP}\) is the brevity penalty to account for shorter candidate sequences. BLEU captures overall fluency but may be diluted by tokens outside the masked API elements.

\textbf{CodeBERTScore} CodeBERTScore, inspired by BERTScore~\cite{zhangbertscore}, measures semantic consistency between generated and reference code by computing the cosine similarity of their token representations. Specifically, CodeBERTScore computes the F scores by combining the precision and recall. Precision (\( \text{CodeBERTScore}_P \)) is calculated as the average maximum similarity between each token in the generated code and tokens in the reference code, while recall (\( \text{CodeBERTScore}_R \)) is computed as the average maximum similarity between each token in the reference code and tokens in the generated code. These are combined using a weighted harmonic mean to produce the final score. In this work, we adopt \( \text{CodeBERTScore}_{F_3} \), which prioritizes recall over precision, as it better aligns with functional correctness in code generation tasks. The formula is defined as:

\[ \begin{aligned} \text{CodeBERTScore}_{F_3} &= \\ &\frac{10 \cdot \text{CodeBERTScore}_P \cdot \text{CodeBERTScore}_R} {9 \cdot \text{CodeBERTScore}_P + \text{CodeBERTScore}_R} \end{aligned} \]

While CodeBERTScore is more flexible than EM in capturing semantic similarity, it has limitations. In particular, it may underestimate correctness when two different but functionally equivalent APIs are used, since token-level embeddings may not capture equivalence unless such relationships were observed during pretraining. Thus, CodeBERTScore should be interpreted as an approximation of semantic similarity.

\textbf{Exact Match (EM)} EM is defined as the count of generated code instances \(\hat{Y}\) whose abstract syntax trees (ASTs) exactly match the ASTs of the corresponding reference/ground-truth code \(Y^*\). It is a strict metric that evaluates syntactic equivalence and may exclude reasonable outputs that differ in syntax but are semantically correct. In our study, EM is computed at the level of the masked API element (method name or parameter), rather than the entire snippet, to directly assess correctness of API usage. EM serves as a lower bound, as different programs can achieve the same functionality but be written differently. The formula is defined as:

\[
\texttt{EM}(\hat{Y}, Y^*) \triangleq \sum_{\hat{y} \in \hat{Y}} \texttt{matches}(\texttt{AST}(\hat{y}), \texttt{AST}(y^*))
\]

A higher EM score consistently indicates stronger performance. However, since human-written code is used as ground truth, EM may penalize models that generate different but still valid API usages. We therefore report EM alongside BLEU, CodeBERTScore, and precision/recall to provide a more complete evaluation of API invocation correctness.

\subsection{EM Failure Qualitative Example Analysis}
\label{subsec:failure}

We examined samples that failed on the EM score, indicating that they are highly likely to be misused by LLMs.  We employed an iterative qualitative analysis approach inspired by grounded theory and qualitative content analysis. 
We randomly sampled 300 pairs of failed cases from each model, categorized by element type (method or parameter) and programming language (Python or Java), yielding a total of 1,200 failure samples. This sample size ensures statistical significance, achieving a 99\% confidence level with a 5\% margin of error.\footnote{\url{https://www.surveysystem.com/sscalc.htm}}

Two authors, each with over five years of programming experience in Python and Java, reviewed and categorized the API misuse samples. We adopt a closed coding strategy by mapping LLM-generated API misuses to Wei et al.’s taxonomy~\cite{wei2024demystifying}, which classifies misuses along two dimensions: violation types (Missing, Redundant, Replacement, Outdated) and API elements (Method, Parameter, Condition). However, because our evaluation is based on isolated code snippets, without access to full projects, temporal version histories, or documentation, we focus only on the violation types that are reliably annotatable in this setting: \textit{Missing} and \textit{Redundant}.

We exclude \textit{Replacement} and \textit{Outdated} categories, as accurately identifying whether a method has been superseded (Outdated) or semantically replaced (Replacement) often requires knowledge of API evolution, project-specific context, or historical intent, which are unavailable in our static snippet-based setting. For example, detecting an outdated method would require knowing the release timeline or deprecation status of the corresponding library, while identifying a replacement misuse would require understanding the intended functionality and its modern equivalent. Since these contextual signals are absent from isolated code samples, including them could lead to annotation ambiguity and unreliable labeling. Consequently, we focus on misuses that can be confidently recognized from self-contained code evidence while maintaining conceptual consistency with Wei et al.’s taxonomy~\cite{wei2024demystifying}.

To analyze LLM-specific characteristics, we annotate two recurring surface patterns, \textit{intent misuse} and \textit{hallucination}, as tags within the broader Missing/Redundant framework. For example, hallucinated methods typically reflect a non-existent method generated in place of a missing or misused one, and can be understood as a form of \textit{Missing Method}. Similarly, intent misuses, while superficially aligned with Replacement–Method, are treated as semantic misfires under \textit{Redundant or Incorrect Method}, when replacement judgment is unreliable. This constrained coding preserves alignment with prior taxonomies while recognizing the limits of static code-only analysis in LLM-generated contexts.

First, the two authors independently familiarized themselves with the dataset by reviewing a subset of API misuse cases. Each annotator labelled 50 randomly selected samples, took detailed notes on their observations, 
During this process, they were allowed to search online for API information when necessary to validate the correctness of the method and parameter usage.
Following this initial review, the annotators conducted axial coding, systematically grouping similar misuses into broader conceptual categories based on shared characteristics. They then met to compare their annotations, resolve discrepancies, and iteratively refine their categorization framework. Through multiple review cycles, they refined their labels, clarified ambiguous definitions, and introduced new categories where necessary. In the end, they progressively consolidated similar misuse patterns into four distinct categories: \textit{Intent Misuse}, \textit{Hallucination}, \textit{Missing Parameter}, \textit{Redundant API Call}. 
With the finalized codebook, the two authors independently labelled the remaining API misuse cases. To assess inter-annotator agreement, we computed Cohen’s Kappa score~\cite{mchugh2012interrater}, achieving a final value of 0.97, indicating almost perfect agreement. The annotation process spanned 300 hours, encompassing multiple iterations of refinement to ensure accuracy and consistency.

It is important to note that while we focus on four main categories of API misuse, there are other potential misuse types, such as security-related or deprecation issues. However, these types are more knowledge-intensive and cannot be easily identified without sufficient contextual information or detailed code snippets. For instance, detecting deprecated methods or security vulnerabilities requires understanding the specific API versions and security considerations, which cannot be reliably determined from code snippets alone. Therefore, our analysis focuses on categories that are more directly identifiable through the provided code and are less dependent on external, domain-specific knowledge. We present each category, which is introduced below, with illustrative examples.

\begin{figure}[!h]
\centering
\begin{lstlisting}[language=diffpy]
 import numpy as np
 from blmath.numerics import vx

 def farthest(from_point, to_points):
     '''
     Find the farthest point among the inputs, to the given point.
     Return a tuple: farthest_point, index_of_farthest_point.
     '''
-    absolute_distances = (*@\color{javared}{\textbf{vs.abs}}@*)(to_points - from_point)
+    absolute_distances = (*@\color{javagreen}{\textbf{vx.magnitude}}@*) (to_points - from_point)
     index_of_farthest_point = np.argmax(absolute_distances)
     farthest_point = to_points[index_of_farthest_point]
     return farthest_point, index_of_farthest_point
\end{lstlisting}

\caption{A diff-formatted example of \textbf{intent misuse}: Incorrect use of \textcolor{javared}{\textbf{\texttt{abs}}} instead of \textcolor{javagreen}{\textbf{\texttt{magnitude}}} in Python.}

\label{fig:intent_misuse_python}
\end{figure}
% \FloatBarrier

\subsection{API Misuse Classification}
\paragraph{LLM-Specific -- Intent Misuse}

\textbf{\textit{Intent misuse occurs when a model selects an API or method that is syntactically correct but does not align with the intended functionality or task.}} This results in incorrect behavior due to the model’s misunderstanding of the method’s purpose. In \autoref{fig:intent_misuse_python}, the method \texttt{farthest()} is intended to compute the farthest point from a reference point, which requires calculating vector magnitudes. The model incorrectly uses \texttt{np.abs}, a \texttt{numpy} function that returns element-wise absolute values, rather than \texttt{vx.magnitude}, which directly computes vector magnitudes. Although syntactically valid, this misuse reflects a semantic mismatch between the selected API and the intended operation. While such cases may be superficially mapped to the Replacement–Method category in Wei et al.’s taxonomy~\cite{wei2024demystifying}, prior studies have not systematically examined intent misuses in the context of LLM-generated code. We adapt and extend the taxonomy to explicitly capture this recurring pattern, which emerges frequently in LLM-assisted code completion. Importantly, we distinguish intent misuses from cases where multiple APIs could plausibly achieve the same functionality (e.g., different logging or plotting methods). In such cases, annotators did not label the output as misuse, since the generated API remained consistent with the overall intent of the code.

\begin{figure}[!h]
\centering
\begin{lstlisting}[language=diffjava]
import java.util.Collections;

public static List<String> getMenuTitles(final Resources res) {
    final List<String> menuList = new ArrayList<String>();
-   strArray = res.getStringArray(R.array.navigation_main_menu_titles);
+   strArray = res.getStringArray(R.array.navigation_main_titles);
    Collections.addAll(menuList,  strArray)
    if (!FeatureConfig.USE_OPENDATA) {
        menuList.remove(MENU_STATISTICS_INDEX);
    }
    return menuList;
}
\end{lstlisting}

\begin{lstlisting}[language=diffjava]
import us.ihmc.commons.MathTools;

@Override
public void initialize() {
    currentTime.set(0.0);
-   MathTools.setCubic(trajectoryTime.getDoubleValue(), 0.0, Double.POSITIVE_INFINITY);
+   MathTools.checkIntervalContains(trajectoryTime.getDoubleValue(), 0.0, Double.POSITIVE_INFINITY);
    parameterPolynomial.setQuintic(0.0, trajectoryTime.getDoubleValue(), 0.0, 0.0, 0.0, 1.0, 0.0, 0.0);

    currentOrientation.set(initialOrientation);
    currentAngularVelocity.setToZero();
    currentAngularAcceleration.setToZero();
}
\end{lstlisting}
\caption{Diff-formatted Examples of \textbf{hallucination misuse}: non-existent \textcolor{javared}{\textbf{\texttt{setCubic}}} instead of \textcolor{javagreen}{\textbf{\texttt{checkIntervalContains}}} in Java.}
\label{fig:hallucination}
\end{figure}
% % \FloatBarrier
\paragraph{LLM-Specific -- Hallucination Misuse} 

\textbf{\textit{Hallucination occurs when a model introduces an entirely incorrect API method or parameter that does not exist or is not required for the task. The model may fabricate an API call that seems plausible but is invalid in context, leading to erroneous behavior or compilation/runtime errors.}} Although prior work has discussed hallucination broadly in LLMs~\cite{liu2024exploring,zhang2024llm,tian2024codehalu}, this work presents the first systematic categorization of hallucination as an API misuse type, where models generate non-existent API methods or parameters. We adapt existing taxonomies by explicitly introducing this category to capture a recurring and distinctive error pattern in LLM-generated code. 
As shown in \autoref{fig:hallucination}, the model incorrectly predicts \texttt{res.getStringArray(R.array.navigation\\\_main\_menu\_titles)} instead of the correct \texttt{res.getStringArray(R.array.navigation\\
\_main\_titles)}. The fabricated parameter name represents a case of \textit{parameter hallucination}. In the same figure, the model generates a non-existent method call \texttt{parameterPolynomial.setCubic(...)} in place of the valid \texttt{MathTools.checkIntervalContains()}, illustrating \textit{method hallucination}. Both cases highlight how LLMs may generate plausible but invalid APIs, a failure mode that is not emphasized in prior human-centric misuse studies but emerges prominently in code completion tasks.

\begin{figure}[!h]
\centering
\begin{lstlisting}[language=diffjava]
import org.chromium.base.PathUtils;

public static void loadLibrary() {
    Context appContext = ContextUtils.getApplicationContext();
-   PathUtils.setPrivateDataDirectorySuffix(PRIVATE_DATA_DIRECTORY_SUFFIX);
+   PathUtils.setPrivateDataDirectorySuffix(PRIVATE_DATA_DIRECTORY_SUFFIX, appContext);

    try {
        LibraryLoader libraryLoader = LibraryLoader.get(LibraryProcessType.PROCESS_WEBVIEW);
        libraryLoader.loadNow(appContext);
        libraryLoader.switchCommandLineForWebView();
    } catch (ProcessInitException e) {
        throw new RuntimeException("Cannot load WebView", e);
    }
}
\end{lstlisting}
\caption{A diff-formatted example of \textbf{missing item misuse}: Omission of the required \textcolor{javared}{\textbf{\texttt{appContext}}} parameter in \textcolor{javagreen}{\textbf{\texttt{setPrivateDataDirectorySuffix}}} in Java.}
\label{fig:java_param_missing_misuse}
\end{figure}
% \FloatBarrier

\paragraph{Generic -- Missing Item Misuse~\cite{wei2024demystifying}} \textbf{\textit{Missing item misuse occurs when the model omits a required API method or parameter, which results in incomplete or incorrect behavior. The model may fail to include a necessary parameter or method, leading to errors or malfunctioning of the system. This type of misuse typically occurs when the model does not fully understand the API's requirements or ignores important details that are essential for correct usage.}} In \autoref{fig:java_param_missing_misuse}, the model incorrectly predicts the API usage by omitting a necessary parameter from the method \texttt{PathUtils.setPrivateDataDirectorySuffix()}. The expected method call should include both the \texttt{PRIVATE\_DATA\_DIRECTORY\_SUFFIX} constant and the \texttt{appContext} parameter. The model fails to include \texttt{appContext}, which is required for the method to properly set the private data directory suffix. This omission leads to parameter omission misuse, where the model overlooks a crucial parameter in the API call, causing potential runtime errors or unexpected behavior.

\begin{figure}[!h]
\centering
\begin{lstlisting}[language=diffjava]
import org.chromium.base.PathUtils;

public static void loadLibrary() {
    Context appContext = ContextUtils.getApplicationContext();
    LibraryLoader libraryLoader = LibraryLoader.get(LibraryProcessType.PROCESS_WEBVIEW);

    try {
        // Redundant method chaining
-       libraryLoader.loadNow(appContext).initialize();  // initialize() is unnecessary
+       libraryLoader.loadNow(appContext);  // Correct: loadNow() is sufficient

        libraryLoader.switchCommandLineForWebView();
    } catch (ProcessInitException e) {
        throw new RuntimeException("Cannot load WebView", e);
    }
}
\end{lstlisting}
\caption{A diff-formatted example of \textbf{redundancy misuse}: Unnecessary chaining of 
\textcolor{javared}{\textbf{\texttt{initialize()}}} after 
\textcolor{javagreen}{\textbf{\texttt{loadNow()}}} in Java. 
According to the official API documentation, \texttt{loadNow()} 
already performs the necessary initialization, so an additional 
call to \texttt{initialize()} has no effect and is therefore redundant.}
\label{fig:java_redundancy_misuse}
\end{figure}
% \FloatBarrier

\paragraph{Generic -- Redundancy~\cite{wei2024demystifying}} \textbf{\textit{Redundancy misuse occurs when the model includes unnecessary method chaining or repeated API calls, which do not contribute to the task and lead to inefficiency or potential errors.}} In the example provided in \autoref{fig:java_redundancy_misuse}, the model incorrectly chains the \texttt{initialize()} method after \texttt{loadNow()}, even though \texttt{loadNow()} already performs the necessary initialization. This redundant method of chaining can lead to confusion and potential side effects. The correct behavior would involve using only the necessary method call, \texttt{loadNow()}, which is sufficient for the task.

\section{Results}
\label{sec:results}
\subsection{RQ1: How well do LLMs handle API invocation in code generation?}

\begin{table}[h!]
\centering
\caption{Performance comparison of StarCoder, Copilot, and Qwen2.5-Coder in Python and Java for method and parameter prediction tasks. EM is reported as percentage over 3{,}000 samples.}
\resizebox{\linewidth}{!}{
\begin{tabular}{ll|ccc|ccc}
\toprule
\multirow{2}{*}{}  & \multirow{2}{*}{\textbf{Model}} & \multicolumn{3}{c|}{\textbf{Python}} & \multicolumn{3}{c}{\textbf{Java}} \\
& & BLEU & CodeBERTScore & EM (\%) & BLEU & CodeBERTScore & EM (\%) \\
\midrule
\multirow{3}{*}{Method} 
    & StarCoder & 84.6 & 92.3 & 75.6 & 81.2 & 86.4 & 44.3 \\
    & Copilot   & 88.3 & 92.6 & 63.7 & 86.1 & 87.7 & 53.1 \\
    & Qwen2.5-Coder & 92.5 & 95.2 & 85.4 & 91.4 & 92.3 & 61.2 \\
\midrule
\multirow{3}{*}{Parameters} 
    & StarCoder & 85.0 & 92.8 & 77.1 & 82.0 & 87.0 & 44.3 \\
    & Copilot   & 89.0 & 93.0 & 65.2 & 87.0 & 88.2 & 53.9 \\
    & Qwen2.5-Coder & 93.4 & 91.1 & 80.2 & 93.6 & 94.7 & 63.5 \\
\bottomrule
\end{tabular}}
\label{tab:model_comparison}
\end{table}

We first wanted to investigate how well our studied LLMs can generate code to correctly invoke APIs in code generation. As described above, we sample 3,000 distinct code snippets in Python and Java, respectively, from The Stack dataset. After prompting StarCoder, Copilot, and Qwen2.5-Coder to complete \textit{Method Prediction} and \textit{Parameter Generation} independently, we collect 36,000 samples in total. We measure the correctness of the API invocation using three metrics introduced in \autoref{subsec:metric}: BLEU, CodeBERTScore, and EM. \autoref{tab:model_comparison} summarizes the performance of the three models on in-the-wild API usage tasks for both Python and Java.

\textbf{For Python, the results reveal distinct trends across the models.}
Copilot and Qwen2.5-Coder achieve higher BLEU (88.3 and 92.5) and CodeBERTScore (92.6 and 95.2) than StarCoder (84.6 and 92.3), indicating that they produce more fluent and semantically consistent completions. However, StarCoder maintains competitive exact match performance (75.6\%), while Qwen2.5-Coder achieves the highest EM (85.4\%), suggesting that it not only generates semantically correct invocations but also matches human-written references more precisely. This improvement likely results from Qwen2.5-Coder’s stronger multilingual code representation and its extensive code-centric training corpus. The divergence between EM and the other metrics highlights how LLMs may generate plausible but syntactically different API invocations, a challenge that remains especially noticeable in Copilot’s outputs.

\textbf{For Java, both Copilot and Qwen2.5-Coder outperform StarCoder across all metrics.}
Copilot achieves an EM of 53.1\%, while Qwen2.5-Coder achieves 61.2\%, compared to StarCoder’s 44.3\%. Similar trends are observed for BLEU and CodeBERTScore, where Qwen2.5-Coder leads (BLEU 91.4, CodeBERTScore 92.3), followed by Copilot (86.1 and 87.7) and StarCoder (81.2 and 86.4). This indicates that Qwen2.5-Coder, although not instruction-tuned, generalizes better to Java APIs, possibly due to broader coverage and better pretraining on strongly typed languages.

The comparison among these three decoder-based models highlights how pretraining strategies and dataset diversity shape model behavior.
StarCoder shows strong memorization and syntactic alignment in Python due to its Python-focused fine-tuning. Copilot achieves consistent semantic fluency across languages through large-scale, multi-language exposure, while Qwen2.5-Coder balances both, delivering strong exact matches and high fluency simultaneously. This suggests that newer open-source models are closing the performance gap with proprietary systems in real-world API invocation tasks.

\begin{figure}[!ht]
\centering
% ---------- Python: Method ----------
\subfloat[Wrong Method Infilling in Python\label{fig:method-prediction-python}]{
\begin{tikzpicture}[scale=0.8]
    \draw[fill=yellow!20,opacity=0.5] (0,0) circle(1.5);    % StarCoder
    \draw[fill=blue!20,opacity=0.5] (1.2,0) circle(1.5);    % Copilot
    \draw[fill=green!20,opacity=0.5] (0.6,1.0) circle(1.5); % Qwen2.5
    % Labels
    \node at (-1.4,-1.6) {\scriptsize StarCoder};
    \node at (2.5,-1.6) {\scriptsize Copilot};
    \node at (0.6,2.8) {\scriptsize Qwen2.5-Coder};
    % Numbers
    \node at (-0.9,-0.3) {126};
    \node at (2.0,-0.3) {466};
    \node at (0.6,1.9) {15};
    \node at (0.6,-0.8) {264};
    \node at (-0.5,0.9) {64};
    \node at (1.6,0.9) {81};
    \node at (0.6,0.3) {278};
\end{tikzpicture}
}
\hfill
% ---------- Python: Parameter ----------
\subfloat[Wrong Parameter Generation in Python\label{fig:param-gen-python}]{
\begin{tikzpicture}[scale=0.8]
    \draw[fill=yellow!20,opacity=0.5] (0,0) circle(1.5);
    \draw[fill=blue!20,opacity=0.5] (1.2,0) circle(1.5);
    \draw[fill=green!20,opacity=0.5] (0.6,1.0) circle(1.5);
    \node at (-1.4,-1.6) {\scriptsize StarCoder};
    \node at (2.5,-1.6) {\scriptsize Copilot};
    \node at (0.6,2.8) {\scriptsize Qwen2.5-Coder};
    \node at (-0.9,-0.3) {138};
    \node at (2.0,-0.3) {430};
    \node at (0.6,1.9) {54};
    \node at (0.6,-0.8) {172};
    \node at (-0.5,0.9) {98};
    \node at (1.6,0.9) {163};
    \node at (0.6,0.3) {279};
\end{tikzpicture}
}

\vspace{1em}

% ---------- Java: Method ----------
\subfloat[Wrong Method Infilling in Java\label{fig:method-prediction-java}]{
\begin{tikzpicture}[scale=0.8]
    \draw[fill=yellow!20,opacity=0.5] (0,0) circle(1.5);
    \draw[fill=blue!20,opacity=0.5] (1.2,0) circle(1.5);
    \draw[fill=green!20,opacity=0.5] (0.6,1.0) circle(1.5);
    \node at (-1.4,-1.6) {\scriptsize StarCoder};
    \node at (2.5,-1.6) {\scriptsize Copilot};
    \node at (0.6,2.8) {\scriptsize Qwen2.5-Coder};
    \node at (-0.9,-0.3) {233};
    \node at (2.0,-0.3) {24};
    \node at (0.6,1.9) {72};
    \node at (0.6,-0.8) {514};
    \node at (-0.5,0.9) {203};
    \node at (1.6,0.9) {148};
    \node at (0.6,0.3) {721};
\end{tikzpicture}
}
\hfill
% ---------- Java: Parameter ----------
\subfloat[Wrong Parameter Generation in Java\label{fig:param-gen-java}]{
\begin{tikzpicture}[scale=0.8]
    \draw[fill=yellow!20,opacity=0.5] (0,0) circle(1.5);
    \draw[fill=blue!20,opacity=0.5] (1.2,0) circle(1.5);
    \draw[fill=green!20,opacity=0.5] (0.6,1.0) circle(1.5);
    \node at (-1.4,-1.6) {\scriptsize StarCoder};
    \node at (2.5,-1.6) {\scriptsize Copilot};
    \node at (0.6,2.8) {\scriptsize Qwen2.5-Coder};
    \node at (-0.9,-0.3) {443};
    \node at (2.0,-0.3) {232};
    \node at (0.6,1.9) {109};
    \node at (0.6,-0.8) {388};
    \node at (-0.5,0.9) {223};
    \node at (1.6,0.9) {146};
    \node at (0.6,0.3) {617};
\end{tikzpicture}
}

\caption{Overlap of incorrect API usages among StarCoder, Copilot, and Qwen2.5-Coder for method infilling and parameter generation in Python and Java.}
\label{fig:api_usage_comparison}
\end{figure}

\autoref{fig:api_usage_comparison} shows the overlap and discrepancies in API misuse among the three models. In Python Method Infilling (\autoref{fig:method-prediction-python}), StarCoder has 126 unique errors, Copilot 466, and Qwen2.5-Coder 15, with substantial overlaps where 278 errors are shared by all three models. Similar patterns hold for Java Method Infilling (\autoref{fig:method-prediction-java}), where StarCoder has 233 unique errors, Copilot 24, and Qwen2.5-Coder 72, again with a large shared subset (721). For parameter generation, the models exhibit analogous overlap patterns where shared errors dominate, although Qwen2.5-Coder consistently produces fewer unique misuses. These results indicate that while Qwen2.5-Coder shows fewer unique errors, all three models converge on many of the same failure cases, reflecting persistent limitations in LLM reasoning about API semantics.
The observed differences can be traced back to each model’s training focus. StarCoder’s specialization in Python contributes to high syntactic fidelity but limited cross-language transfer, Copilot benefits from multi-language fine-tuning but sacrifices exactness, and Qwen2.5-Coder attains strong generalization through diverse code-centric pretraining without instruction tuning.

\begin{findingbox}
\textbf{Finding 1:} Qwen2.5-Coder achieves the highest overall accuracy across both languages, combining strong fluency with precise matching. Copilot follows closely, while StarCoder remains competitive in Python due to domain specialization. Despite these differences, all three models share a substantial number of common API misuses, which shows that correct API invocation remains a fundamental challenge for current LLMs.
\end{findingbox}

\subsection{RQ2: How do LLMs misuse APIs in the wild?}

\begin{table}[ht]
\centering
\caption{Comparison of API Misuse by StarCoder in Java and Python. ``None'' indicates that the LLM-based API usage was considered acceptable based on human inspection.}
\resizebox{\linewidth}{!}{
\begin{tabular}{l|cc|cc}
\toprule
 \multirow{2}{*}{\textbf{Misuse Type}} & \multicolumn{2}{c|}{\textbf{Java}} & \multicolumn{2}{c}{\textbf{Python}} \\
 & \textbf{Params (\%)} & \textbf{Methods (\%)} & \textbf{Params (\%)} & \textbf{Methods (\%)} \\
\midrule
      Intent             & 5.5  & 27.0 & 7.4  & 30.4 \\
  Hallucination          & 43.3 & 34.6 & 31.3 & 40.9 \\
     Redundancy          & 8.2  & 10.2 & 14.8 & 5.3  \\
     Missing Item        & 15.0 & 11.3 & 26.5 & 3.0  \\\midrule
          None           & 28.0 & 16.9 & 20.0 & 20.4 \\
\bottomrule
\end{tabular}}
\label{tab:api_misuse_comparison_starcoder}
\end{table}

\begin{table}[ht]
\centering
\caption{Comparison of API Misuse by Copilot in Java and Python. ``None'' indicates that the LLM-based API usage was considered acceptable based on human inspection.}
\resizebox{\linewidth}{!}{
\begin{tabular}{l|cc|cc}
\toprule
 \multirow{2}{*}{\textbf{Misuse Type}} & \multicolumn{2}{c|}{\textbf{Java}} & \multicolumn{2}{c}{\textbf{Python}} \\
 & \textbf{Params (\%)} & \textbf{Methods (\%)} & \textbf{Params (\%)} & \textbf{Methods (\%)} \\
\midrule
      Intent             & 4.0  & 22.0 & 6.0  & 25.0 \\
  Hallucination          & 39.0 & 29.0 & 25.0 & 33.0 \\
     Redundancy          & 7.0  & 12.0 & 12.0 & 7.0  \\
     Missing Item        & 15.0 & 12.5 & 29.0 & 7.7  \\\midrule
          None           & 35.0 & 24.5 & 28.0 & 27.3 \\
\bottomrule
\end{tabular}}
\label{tab:api_misuse_comparison_copilot}
\end{table}

\begin{table}[ht]
\centering
\caption{Comparison of API Misuse by Qwen2.5-Coder in Java and Python. ``None'' indicates that the LLM-based API usage was considered acceptable based on human inspection.}
\resizebox{\linewidth}{!}{
\begin{tabular}{l|cc|cc}
\toprule
 \multirow{2}{*}{\textbf{Misuse Type}} & \multicolumn{2}{c|}{\textbf{Java}} & \multicolumn{2}{c}{\textbf{Python}} \\
 & \textbf{Params (\%)} & \textbf{Methods (\%)} & \textbf{Params (\%)} & \textbf{Methods (\%)} \\
\midrule
      Intent             & 4.2  & 20.5 & 6.5  & 23.6 \\
  Hallucination          & 33.5 & 26.8 & 23.0 & 29.4 \\
     Redundancy          & 6.3  & 10.5 & 10.7 & 6.4  \\
     Missing Item        & 13.0 & 11.7 & 25.4 & 8.1  \\\midrule
          None           & 43.0 & 30.5 & 34.4 & 32.5 \\
\bottomrule
\end{tabular}}
\label{tab:api_misuse_comparison_qwen}
\end{table}

We further analyze the patterns and distributions of API misuse across StarCoder, Copilot, and Qwen2.5-Coder. Specifically, we study the annotated outputs introduced in \autoref{subsec:failure}, which correspond to samples that failed on EM score. \autoref{tab:api_misuse_comparison_starcoder}, \autoref{tab:api_misuse_comparison_copilot}, and \autoref{tab:api_misuse_comparison_qwen} show the distribution of API misuse types across Java and Python, categorized by methods and parameters.

\textbf{Hallucination remains the most frequent misuse across all three models, although its prevalence varies.} For StarCoder, hallucinations dominate all configurations, accounting for up to 43.3\% of parameter misuses in Java. Copilot shows slightly reduced rates, while Qwen2.5-Coder exhibits further decreases, with hallucination rates below 30\% in most cases. This indicates that Qwen2.5-Coder is better grounded in valid API namespaces, reducing the tendency to generate non-existent or version-inconsistent APIs.

\textbf{Missing Item misuses are prominent for Python parameters, particularly for Copilot and Qwen2.5-Coder.} Copilot omits required parameters in 29.0\% of Python parameter cases, making it the leading error type for that configuration. Qwen2.5-Coder also shows relatively high omission rates (25.4\%), although slightly lower than Copilot’s. This pattern suggests that the challenge of inferring correct argument lists persists even for newer models, especially in Python where argument flexibility and version heterogeneity increase complexity.

\textbf{Intent misuses continue to appear frequently but are less severe in Qwen2.5-Coder.} In Python, StarCoder records 30.4\% intent errors for methods, while Copilot and Qwen2.5-Coder reduce these to 25.0\% and 23.6\% respectively. Similar trends appear in Java. These improvements suggest that newer models capture functional intent more reliably, although confusion between similar API methods still occurs.

\textbf{Redundancy misuses are less frequent but remain consistent across models.} Qwen2.5-Coder reduces redundant arguments and superfluous method calls compared to StarCoder and Copilot, showing the lowest rates overall (around 6–10\%). This aligns with its improved parameter consistency and contextual inference capabilities.

\textbf{Overall, Python parameter completion remains the most error-prone scenario.} The results across all three models indicate that Python’s dynamic typing and overlapping library interfaces still pose major challenges. Frequent hallucination and omission misuses demonstrate that LLMs struggle to fully align API invocations with real-world library constraints, especially for parameters that require contextual understanding of object states or external imports.

\begin{findingbox}
\textbf{Finding 2:} Hallucination remains the most common misuse overall, although Qwen2.5-Coder reduces its occurrence compared with StarCoder and Copilot. Missing Item errors remain prominent in Python parameter prediction for all models. Intent misuses are frequent in method prediction, but Qwen2.5-Coder demonstrates improved grounding and fewer redundant errors, suggesting gradual progress toward more reliable API usage.
\end{findingbox}

\begin{figure*} 
\centering 
\includegraphics[width=\textwidth]{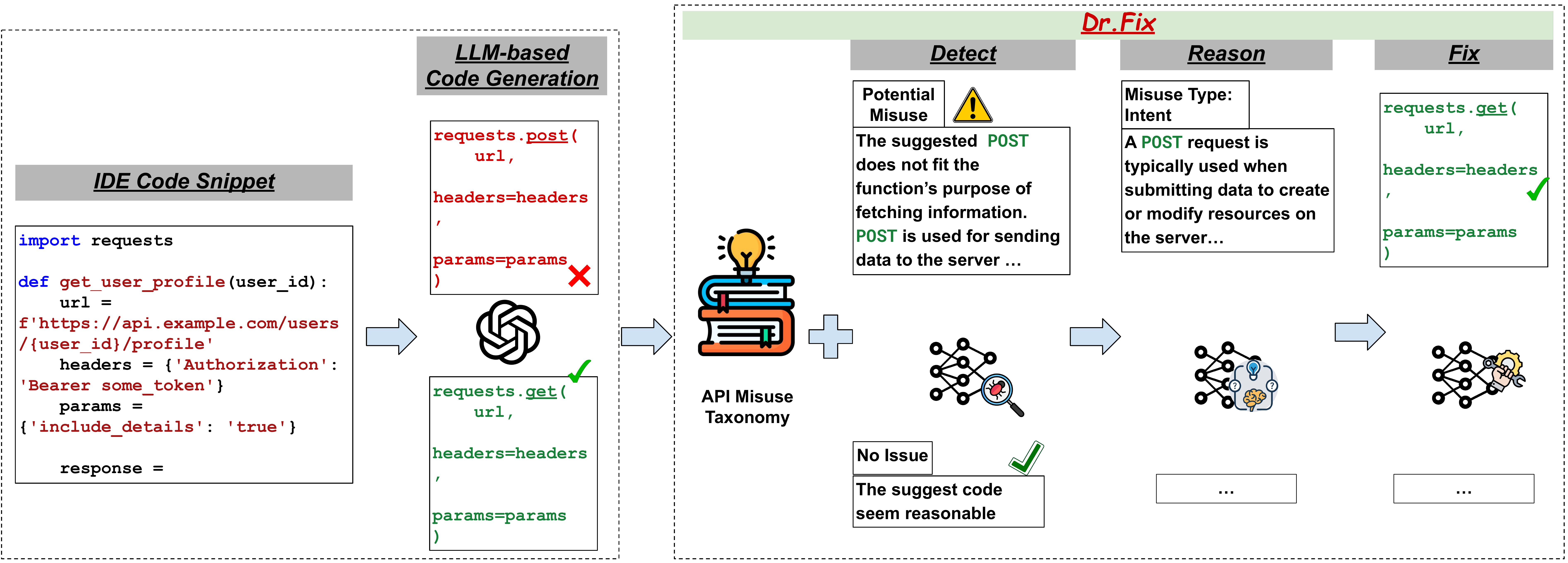} 
\caption{Workflow of Dr.Fix to repair the API misuse.} 
\label{fig:drfix} 
\end{figure*}

\section{Mitigation}
\label{sec:mitigation}
Mitigating LLM-based API misuse in real-world applications presents several challenges. Existing automatic misuse detection methods often require substantial manual effort to apply fixes, underscoring the need for approaches that can both detect and repair misuses. However, program repair methods tailored specifically for API misuse remain largely unexplored. The only relevant work is \cite{zhang2023evaluating}, which evaluated pre-trained models fine-tuned on large-scale git commit patches, but focused only on Java. More recently, Wei et al. proposed LLMAPIDet~\cite{wei2024demystifying}, a detector and repair tool for deep learning libraries based on ChatGPT. While effective in that domain, it is limited by its reliance on closed-source models and its focus on TensorFlow and PyTorch APIs. By contrast, Dr.Fix is designed to be backbone-agnostic and taxonomy-driven, making it applicable to general-purpose third-party APIs across different programming languages.

Repairing LLM-generated API misuses is particularly important because such misuses often arise from the way LLMs generate code, rather than from human misunderstanding of documentation. Unlike traditional APR methods, which rely on static analysis or heuristics, LLM-based repair can leverage contextual knowledge of APIs and natural language reasoning to propose fixes. This enables handling of misuse types such as hallucinations or incorrect intent selections, which are less common in human-written code but prevalent in LLM outputs. At the same time, it is valuable to investigate whether LLMs reproduce the same kinds of misuses as humans, or if they introduce distinctive patterns that call for new repair strategies.

We explore a promising direction for API misuse repair: prompt-based LLM program repair. Unlike conventional APR methods trained on task-specific datasets, prompt-based approaches offer greater flexibility and generalization. Prior work has shown that LLMs are capable of self-repair and debugging through structured prompting~\cite{chenteaching}, motivating our design of Dr.Fix.

\subsection{Dr.Fix: Repairing API Misuse via Detection and Reasoning}

Repairing LLM-based API misuses requires nuanced semantic understanding, particularly of the implicit intents within the code context. While LLMs can generate repairs when directly prompted, prior work shows that naive prompting often produces syntactically valid but semantically incomplete or inconsistent fixes~\cite{chenteaching,sclarquantifying,miaoselfcheck,mizrahi2024state}. Moreover, recent studies on domain-specific APIs highlight that LLM-based repair is sensitive to how misuses are detected and framed, and simple prompts may overlook subtle but critical constraints~\cite{wei2024demystifying,kwon2023exploring}. These observations motivate the need for structured prompting strategies that incorporate detection, classification, and reasoning, rather than relying on single-pass repair. At the same time, traditional APR methods, reliant on static analysis or heuristic rules, remain limited when applied across diverse programming environments.  

\autoref{fig:drfix} overviews our Dr.Fix approach, which integrates detection and reasoning capabilities through multiple stages of LLM prompting. The process begins with the detection stage, where the model is prompted to recognize potential API misuse within the code. Once potential misuses are identified, a taxonomy classification stage follows, mapping them into predefined categories such as intent or hallucination misuse. In the reasoning stage, the model validates the detected misuse with a detailed analysis that considers the broader context and intended functionality. This stage is motivated by self-verification reasoning~\cite{weng2023large}, which has been effective in natural language processing tasks. Finally, the repair suggestion stage produces candidate fixes that aim to be both syntactically correct and semantically meaningful, aligning with the developer’s intent.  

Incorporating Dr.Fix into the development workflow reduces the manual effort required to address API misuses. By automating both detection and repair, Dr.Fix improves the reliability and efficiency of software development. Our design demonstrates the potential of combining LLM capabilities with structured reasoning techniques to address complex challenges in automated program repair.

\subsection{Backbone Instruction-tuned Models}

LLMs serve distinct purposes in software engineering workflows. 
Models designed for \textit{IDE-based code completion} prioritize efficiency and responsiveness, enabling seamless interaction within development environments. 
In contrast, models aimed at \textit{code repair} focus on semantic correctness and reliability rather than generation speed, as repair involves complex reasoning over code context and intent. 
\autoref{tab:model_comparison} summarizes the speed and cost characteristics of representative models in both categories. 
While the IDE-based code completion results are derived from our previous experiments, in this work we concentrate on LLMs for API misuse repair when evaluating the performance of Dr.Fix. 
To ensure the highest possible repair quality, we adopt frontier LLMs which better represent the current technological upper bound in reasoning and code understanding. 
This distinction allows us to balance efficiency and precision while analyzing the contribution of repair-oriented LLMs to Dr.Fix’s overall effectiveness.

\begin{table}[h!]
\centering
\caption{Model Comparison: Speed and Cost. Speed and cost for all models except StarCoder and Copilot are obtained from OpenRouter~\cite{openrouter}. 
For StarCoder and Copilot, we measure inference speed locally; both are free to use in this setup. We note that the cost is measured per 1M tokens.}
\label{tab:model_comparison}
\resizebox{\linewidth}{!}{%
\begin{tabular}{lccc}
\toprule
\multicolumn{4}{c}{\textit{IDE}} \\
\midrule
\textbf{Metric} & \textbf{StarCoder} & \textbf{Copilot} & \textbf{Qwen2.5-Coder} \\
\midrule
Speed (tokens/s) & 217.4 & 135.0 & 229.0 \\
Cost (\$ in/out) & 0.0 / 0.0 & 0.0 / 0.0 & 0.03 / 0.09 \\
\midrule
\multicolumn{4}{c}{\textit{Repair}} \\
\midrule
\textbf{Metric} & \textbf{Llama 3.1} & \textbf{Qwen2.5-Coder-Instruct} & \textbf{GPT-4o} \\
\midrule
Speed (tokens/s) & 22.65 & 38.17 & 59.3 \\
Cost (\$ in/out) & 0.4 / 0.4 & 0.04 / 0.16 & 2.5 / 10 \\
\bottomrule
\end{tabular}%
}
\end{table}

\paragraph{Llama-3.1} Llama 3.1~\cite{dubey2024llama} is a collection of multilingual LLMs, a collection of pre-trained and instruction-tuned generative models in 8B, 70B, and 405B sizes released by Meta AI. The Llama 3.1 instruction-tuned text-only models (8B, 70B, 405B) are optimized for multilingual dialogue use cases and outperform many available open-source and closed-chat models on common industry benchmarks. Compared to the original Llama 3 series, Llama-3.1 is significantly improved in terms of math and reasoning capabilities. In our evaluation, we use Llama-3.1 70B, a variant achieving state-of-the-art (SOTA) performance on various benchmarks.

\paragraph{Qwen2.5-Coder-Instruct} Qwen2.5-Coder-Instruct~\cite{hui2024qwen2} is a series of code-specific generative models designed to excel in a wide range of programming tasks such as code generation, completion, reasoning, and repair. Built upon the Qwen2.5 architecture, these models come in sizes ranging from 0.5B to 32B parameters and have been pre-trained on a vast corpus of over 5.5 trillion tokens. Through a combination of extensive data cleaning, scalable synthetic data generation, and balanced data mixing, Qwen2.5-Coder-Instruct achieves SOTA performance across more than 10 code-related benchmarks. It consistently outperforms other models of similar size, showcasing impressive capabilities in code intelligence while retaining strong general and math skills. With its permissive licensing, Qwen2.5-Coder-Instruct is positioned to drive innovation in code-related AI research and be widely adopted by developers in real-world applications. We chose Qwen2.5-Coder-Instruct 32B, the most capable one among the series.

\paragraph{GPT-4o} GPT-4o\footnote{\url{https://openai.com/index/hello-gpt-4o/}} is a highly advanced variant of OpenAI's GPT-4 model, fine-tuned specifically for enhanced efficiency and specialized performance across various natural language processing and multimodal tasks. Available in different configurations, GPT-4o delivers superior reasoning, problem-solving, and contextual understanding capabilities compared to its predecessors. With a focus on optimizing performance while reducing computational overhead, GPT-4o excels in both complex language tasks and more specialized use cases, such as coding, translation, and content generation. Leveraging a blend of advanced training techniques, including reinforcement learning from human feedback, GPT-4o sets a new standard for AI models in terms of accuracy, responsiveness, and versatility, making it ideal for real-world applications across industries.

\subsection{Hyperparameters}

In this study, we set a maximum token length to 2048, a temperature of 0, and a top-p of 0.95. The few-shot learning approach~\cite{brown2020language} is used to prompt the models across four stages: detection, classification, reasoning, and repair suggestion. Specifically, we use two shots per stage as a demonstration to ensure the output format. For each type inside the API misuse taxonomy, we sample two failure cases to teach the models.

The prompts for each stage are designed as follows: (1) Detect - ``Identify any incorrect usages in the following code snippet based on one of the following categories: Intent, Hallucination, Redundancy, Missing.'' The definition of each category is provided after the instructions. To prevent LLMs from hallucinating API misuse, we append ``Please answer with "No Issue" if there is no obvious API misuse.'' at the end of the prompt. (2) Reason - ``Think step by step and explain why this API usage is incorrect and how it deviates from the intended usage.'' (3) Fix - ``Propose a correction that aligns with the intended functionality of the API.''

The models' performance is evaluated based on BLEU, CodeBERTScore, EM accuracy, and refusal rate to assess their ability to detect and repair API misuses across different categories and programming languages.

The refusal rate is introduced as an additional evaluation metric inspired by \cite{zhang2023survey,xu2024rejection,zhang2024r}. It measures the model's ability to avoid hallucinating an API misuse when none exists in the code snippet. This score is defined as the percentage of instances where the model correctly refuses to detect a misuse when no misuse is present in the code. The formula for the refusal rate is:

\[
\text{Refusal Rate} = \frac{\text{Number of Correct Rejections}}{\text{Total Number of No Misuse Samples}} \times 100
\]

\subsection{Previous Approaches}

We use \cite{zhang2023evaluating} as a representative baseline of learning-aided automated program repair (APR), which outperforms previous APR methods~\cite{chen2019sequencer, zhu2021syntax}.
Zhang et al.\cite{zhang2023evaluating} systematically evaluated pre-trained models, including CodeBERT and CodeT5, on repairing 
API- and semantic-related bugs. Their approach fine-tunes PLMs on a large-scale dataset of human-written 
\texttt{git commit} patches and evaluates repair performance primarily on Java projects. 

This work is relevant as it demonstrates how supervised PLMs can be adapted for code repair. However, it has two key limitations: 
(1) it focuses exclusively on Java and deep learning library APIs, which restricts its applicability to other programming 
languages such as Python, and (2) it depends on fine-tuning closed or specialized models, limiting generalization to 
diverse libraries and cross-language settings. In contrast, our work investigates prompt-based, instruction-tuned 
LLMs (Dr.Fix) that require no fine-tuning, operate on a general API misuse taxonomy, and extend beyond a single domain.  

For comparison, we replicate their best-performing baselines, CodeBERT and CodeT5, for the Java Method and 
Parameter tasks, and report them alongside Dr.Fix in our evaluation.

\begin{table}[ht]
\centering
\caption{10\% Sample of Misuse Types for Java and Python}

\resizebox{\linewidth}{!}{
\begin{tabular}{l|cc|cc}
\toprule
 \multirow{2}{*}{\textbf{Misuse Type}} & \multicolumn{2}{c|}{\textbf{Java}} & \multicolumn{2}{c}{\textbf{Python}} \\
 & \textbf{Params \#} & \textbf{Methods \#} & \textbf{Params \#} & \textbf{Methods \#} \\
\midrule
 Intent                            & 11   & 55   & 10   & 39  \\
Hallucination                         & 89   & 71   & 40   & 53  \\
Redundancy                          & 17   & 21   & 19   & 7   \\
Missing                            & 31   & 23   & 34   & 4   \\
\midrule
\textbf{Total}   & 148 & 170 & 103 & 103 \\
\bottomrule
\end{tabular}}
\label{tab:sample_misuse}
\end{table}

\subsection{Dataset}

We sampled 10\% of code snippets from each annotated misuse type of StarCoder and Copilot in \autoref{sec:case_study}. As shown in \autoref{tab:sample_misuse}, the dataset includes a representative sample of API misuse types across Java and Python. Specifically, the dataset captures the following misuse types: Intent, Hallucination, Redundancy, and Missing. For each type, a proportional number of samples was extracted based on their original distribution. To further enhance the evaluation of repair approaches, including Dr.Fix and other baselines, we introduced an additional 100 snippets that exactly match the reference source code. This addition allows us to test the ability of repair tools to avoid unnecessary hallucinations or misinterpretations of correct API usage. By including these reference-aligned snippets, we can measure the precision of tools in distinguishing between actual misuse and correct implementations.

\subsection{Results Analysis}

\begin{table}[h!]
\centering
\caption{Performance of Models across Different Methods. \textbf{Dr.Fix (w/o taxonomy)} means that the models are not prompted with the defined API misuse taxonomy and demonstrated examples, and perform the detection based on their own understanding. \textbf{Baseline} denotes the LLM-based APR method where the models are prompted to generate the fixed code snippet if they detect the API misuse inside the given snippet.}

\resizebox{!}{0.8\linewidth}{
\begin{tabular}{llcccc}
\toprule
\textbf{Model}   & \textbf{Method}   & \textbf{BLEU} & \textbf{CodeBERTScore} & \textbf{EM (\#)} \\ 
\midrule

\multicolumn{6}{c}{\textit{Java -- Method}}
\\\midrule
\cellcolor{gray!10} CodeBERT  & \cite{zhang2023evaluating} & 12.4 & 15.8 & 3 \\
\midrule
\cellcolor{gray!10} CodeT5   & \cite{zhang2023evaluating} & 20.1 & 24.7 & 5 \\
\midrule
\multirow{3}{*}{Llama-3.1} & Baseline       & 30.5 & 25.7 & 35 \\ 
                       & Dr.Fix (w/o taxonomy) & 65.2 & 66.3 & 75 \\ 
                       & \cellcolor{blue!10}Dr.Fix        & \cellcolor{blue!10}68.9 & \cellcolor{blue!10}69.2 & \cellcolor{blue!10}80 \\ 
\midrule
\multirow{3}{*}{Qwen2.5-Coder-Instruct}  & Baseline       & 36.3 & 28.4 & 24 \\ 
                       & Dr.Fix (w/o taxonomy) & 55.1 & 60.2 & 92 \\ 
                       & \cellcolor{blue!10}Dr.Fix        & \cellcolor{blue!10}58.3 & \cellcolor{blue!10}62.6 & \cellcolor{blue!10}96\\ 
\midrule
\multirow{3}{*}{GPT-4o}  & Baseline       & 57.8 & 46.1 & 52 \\ 
                       & Dr.Fix (w/o taxonomy) & 76.1 & 65.3 & 92 \\ 
                       & \cellcolor{blue!10}Dr.Fix        & \cellcolor{blue!10}79.4 & \cellcolor{blue!10}68.2 & \cellcolor{blue!10}96 \\ 
\midrule
\multicolumn{6}{c}{\textit{Java -- Parameter}}
\\\midrule
\cellcolor{gray!10} CodeBERT  & \cite{zhang2023evaluating} & 13.5 & 17.2 & 2 \\
\midrule
\cellcolor{gray!10} CodeT5    & \cite{zhang2023evaluating} & 22.8 & 26.4 & 6 \\
\midrule
\multirow{3}{*}{Llama-3.1} & Baseline       & 53.6 & 64.6 & 34 \\ 
                       & Dr.Fix (w/o taxonomy) & 72.3 & 85.1 & 72 \\ 
                       & \cellcolor{blue!10}Dr.Fix        & \cellcolor{blue!10}75.8 & \cellcolor{blue!10}88.3 & \cellcolor{blue!10}78  \\ 
\midrule
\multirow{3}{*}{Qwen2.5-Coder-Instruct}  & Baseline       & 55.7 & 73.3 & 21 \\ 
                       & Dr.Fix (w/o taxonomy) & 73.2 & 89.5 & 80 \\ 
                       & \cellcolor{blue!10}Dr.Fix        & \cellcolor{blue!10}76.4 & \cellcolor{blue!10}92.3 & \cellcolor{blue!10}85 \\ 
\midrule
\multirow{3}{*}{GPT-4o} & Baseline       & 62.7 & 82.6 & 26 \\ 
                       & Dr.Fix (w/o taxonomy) & 80.1 & 89.2 & 84 \\ 
                       & \cellcolor{blue!10}Dr.Fix        & \cellcolor{blue!10}82.5 & \cellcolor{blue!10}91.5 & \cellcolor{blue!10}88 \\
\midrule
\multicolumn{6}{c}{\textit{Python -- Method}}
\\\midrule
\multirow{3}{*}{Llama-3.1} & Baseline       & 34.6 & 58.4 & 25\\ 
                       & Dr.Fix (w/o taxonomy) & 54.2 & 87.6 & 55\\ 
                       & \cellcolor{blue!10}Dr.Fix        & \cellcolor{blue!10}57.8 & \cellcolor{blue!10}90.3 & \cellcolor{blue!10}60\\ 
\midrule
\multirow{3}{*}{Qwen2.5-Coder-Instruct}  & Baseline       & 15.1 & 43.8 & 14\\ 
                       & Dr.Fix (w/o taxonomy) & 49.3 & 81.2 & 48 \\ 
                       & \cellcolor{blue!10}Dr.Fix        & \cellcolor{blue!10}52.6 & \cellcolor{blue!10}84.6 & \cellcolor{blue!10}52 \\ 
\midrule
\multirow{3}{*}{GPT-4o}  & Baseline       & 47.0 & 69.5 & 31\\ 
                       & Dr.Fix (w/o taxonomy) & 60.5 & 90.1 & 66\\ 
                       & \cellcolor{blue!10}Dr.Fix        & \cellcolor{blue!10}63.7 & \cellcolor{blue!10}92.7 & \cellcolor{blue!10}71\\ 
\midrule
\multicolumn{6}{c}{\textit{Python -- Parameter}}
\\\midrule
\multirow{3}{*}{Llama-3.1} & Baseline       & 78.7 & 75.7 & 12 \\ 
                       & Dr.Fix (w/o taxonomy) & 82.1 & 84.3 & 40 \\ 
                       & \cellcolor{blue!10}Dr.Fix        & \cellcolor{blue!10}84.5 & \cellcolor{blue!10}86.8 & \cellcolor{blue!10}45 \\ 
\midrule
\multirow{3}{*}{Qwen2.5-Coder-Instruct}  & Baseline       & 81.2 & 62.3 & 10 \\ 
                       & Dr.Fix (w/o taxonomy) & 85.1 & 87.2 & 52  \\ 
                       & \cellcolor{blue!10}Dr.Fix        & \cellcolor{blue!10}87.4 & \cellcolor{blue!10}89.4 & \cellcolor{blue!10}56  \\ 
\midrule
\multirow{3}{*}{GPT-4o}  & Baseline       & 84.8 & 67.2 & 11  \\ 
                       & Dr.Fix (w/o taxonomy) & 86.9 & 90.8 & 34 \\ 
                       & \cellcolor{blue!10}Dr.Fix        & \cellcolor{blue!10}88.6 & \cellcolor{blue!10}92.4 & \cellcolor{blue!10}38 \\ 
\bottomrule
\end{tabular}}
\label{tab:mitigation}
\end{table}
% % \FloatBarrier

The performance of the models across different methods is summarized in \autoref{tab:mitigation}. The results highlight the significant improvement in API misuse detection and repair capabilities achieved by incorporating the Dr.Fix methodology, with and without the taxonomy. This section provides an in-depth analysis of the findings for both Java and Python across the evaluated metrics.

\textbf{Java Performance:} 
For Java, \textbf{both Dr.Fix and Dr.Fix (w/o taxonomy) outperform the baseline method across all metrics}. 
In the \textit{Method} misuse category, GPT-4o with Dr.Fix achieves the highest BLEU score and the highest EM (tied with Qwen2.5-Coder-Instruct), 
demonstrating its ability to generate syntactically and semantically appropriate repairs. The intermediate version, 
Dr.Fix (w/o taxonomy), also shows substantial improvements, with GPT-4o achieving a BLEU score of 76.1 and an EM of 92, 
which further enhance to 79.4 and 96, respectively, with the full taxonomy. Similarly, Qwen2.5-Coder-Instruct and Llama-3.1 
show marked improvements with both versions of Dr.Fix, with BLEU scores increasing from 36.3 to 58.3 and from 30.5 
to 68.9, respectively. In the \textit{Parameter} misuse category, the results follow a similar trend. GPT-4o achieves 
the best BLEU and EM, while Qwen2.5-Coder-Instruct attains the highest CodeBERTScore. Dr.Fix (w/o taxonomy) notably enhances the performance of all 
models, with improvements in both syntactic and semantic alignment, as reflected in the BLEU and CodeBERTScore 
metrics, which are further boosted by the full taxonomy. 

In addition, we include Zhang et al.~\cite{zhang2023evaluating} as representative learning-based APR baselines. Their performance on both Java \textit{Method} and \textit{Parameter} repair is substantially lower than instruction-tuned LLMs. For instance, CodeBERT and CodeT5 achieve BLEU scores of 12.4 and 20.1 on the \textit{Method} task, and 13.5 and 22.8 on the \textit{Parameter} task, respectively, which are far below GPT-4o with Dr.Fix (79.4 and 82.5). This highlights the advantage of instruction-tuned LLMs, especially when guided by structured misuse taxonomies. This gap highlights 
that fine-tuned PLMs trained on commit-level patches are far less effective for API misuse repair than prompt-based 
LLMs. Moreover, these approaches lack awareness of the distinctive misuse patterns emerging in LLM-generated code, 
such as hallucination and intent misuses, limiting their applicability in this setting. These results further confirm 
that Dr.Fix offers not only relative gains over naive prompting baselines, but also a clear advantage over prior 
supervised APR approaches.

\textbf{Python Performance:}
For Python, \textbf{both Dr.Fix and Dr.Fix (w/o taxonomy) demonstrate significant improvements over the baseline method}. In the \textit{Method} misuse category, GPT-4o achieves the highest BLEU score and CodeBERTScore, indicating superior generalization and repair accuracy. The EM score also improves from 31 to 71 with the full taxonomy, while Dr.Fix (w/o taxonomy) achieves an EM of 66. Similarly, Qwen2.5-Coder-Instruct and Llama-3.1 achieve notable gains in BLEU and CodeBERTScore metrics with both versions of Dr.Fix, showcasing the effectiveness of the methodology in improving code repair suggestions. In the \textit{Parameter} misuse category, all models exhibit enhanced performance with Dr.Fix. GPT-4o achieves a BLEU score of 88.6 and a CodeBERTScore of 92.4 with the full taxonomy, surpassing other models in both accuracy and precision. While the baseline method performs reasonably well in this category, both versions of Dr.Fix provide a clear advantage, ensuring that repair suggestions align more closely with developer intentions, with the full taxonomy offering the most significant improvements.

\begin{figure}[h!]
    \centering
    \includegraphics[width=\linewidth]{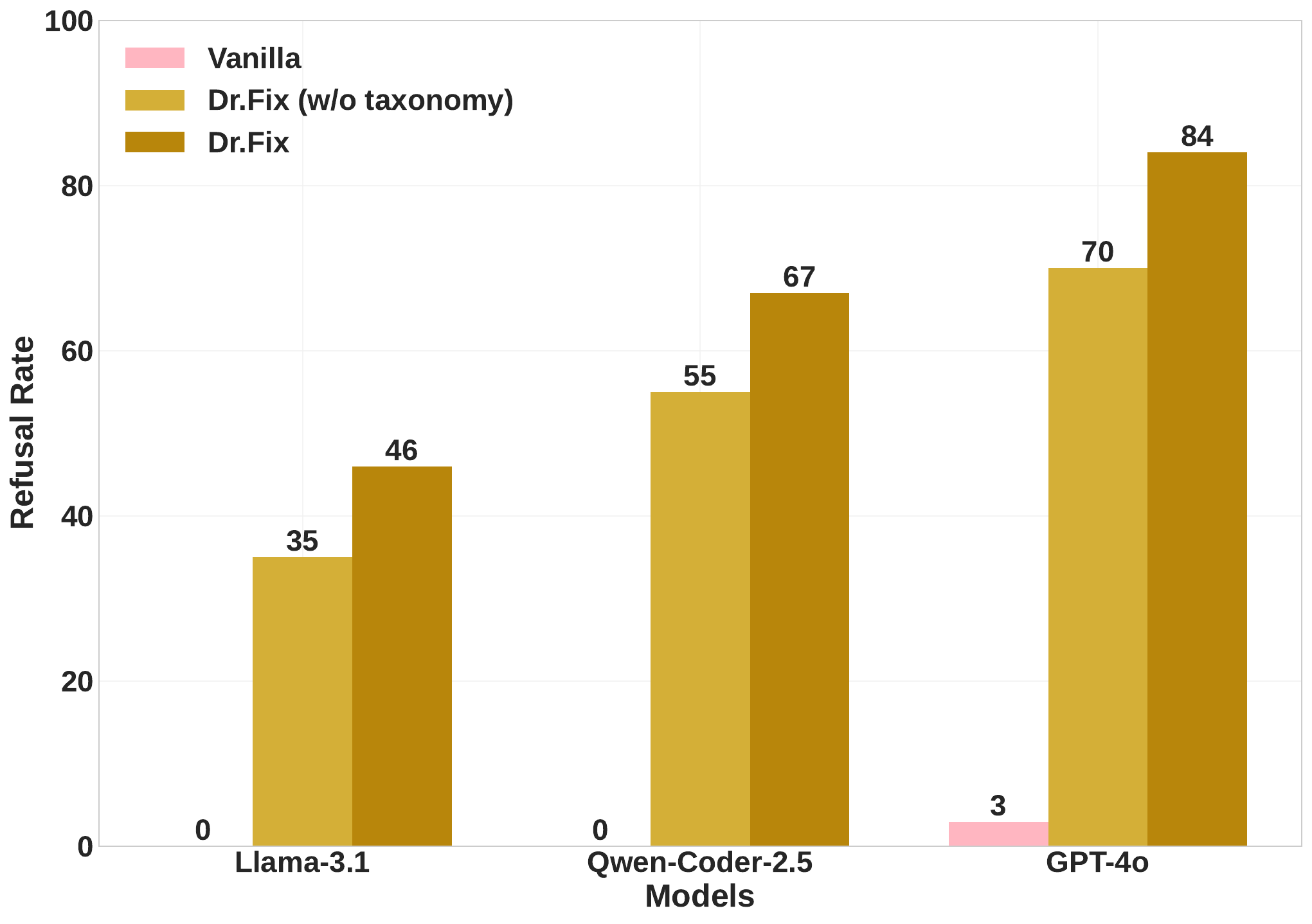}
\caption{Refusal rate comparison of Llama-3.1, Qwen2.5-Coder-Instruct, and GPT-4o with and without Dr.Fix.}
    \label{fig:refusal-rate}
\end{figure}

\textbf{Refusal Rate: }
The refusal rate metric evaluates the models' ability to avoid hallucinating API misuses when no misuse exists. As shown in \autoref{fig:refusal-rate}, \textbf{Dr.Fix significantly enhances the refusal rates across all tested models}. GPT-4o achieved the highest refusal rate of 84\% with Dr.Fix, demonstrating its exceptional ability to differentiate between correct and incorrect API usage. Similarly, Qwen2.5-Coder-Instruct and Llama-3.1 showed substantial improvements, achieving refusal rates of 67\% and 46\%, respectively, compared to their baseline configurations, which had refusal rates of 3\%, 0\%, and 0\%, respectively. Dr.Fix (w/o taxonomy) also showed promising results, with refusal rates of 70\%, 55\%, and 35\% for GPT-4o, Qwen2.5-Coder-Instruct, and Llama-3.1, respectively. Notably, Dr.Fix slightly decreases the refusal rate when no API misuse taxonomy is given for the model, suggesting that it effectively reduces false positives by avoiding unnecessary refusals for correct API usage. These results highlight Dr.Fix's effectiveness in reducing hallucinated API misuses across diverse LLM architectures, with the full taxonomy version providing the most significant improvements.
\begin{findingbox}
Our experimental results demonstrate that Dr.Fix significantly enhances API misuse repair across all models, yielding up to 72 more exact matches than baseline prompting while maintaining high code quality. It also raises refusal rates on misuse-free code by 46--81 percentage points over vanilla prompting (and by 11--14 points over Dr.Fix without the taxonomy) across different architectures, showcasing its effectiveness as a reliable program repair solution for diverse programming languages.
\end{findingbox}

\section{Threats to Validity}
\label{sec:threat}
\subsection{Internal Validity}
One potential threat to internal validity is the reliance on subjective human annotations for the initial classification of API misuse. Although two annotators with over five years of programming experience independently reviewed the samples, biases in understanding or interpretation of the API misuse may have influenced the categorization. We mitigated this by ensuring high inter-annotator agreement, as indicated by the high Cohen’s Kappa score, and resolving disagreements through detailed discussions. Another threat arises from the inherent variability in LLM outputs due to stochastic sampling. To address this, we kept the model parameters fixed across experiments, varying only the prompts as necessary. However, given resource constraints, experiments were conducted once, and results may vary under repeated evaluations. Future studies should consider multiple iterations to assess consistency and variability in repair suggestions. Finally, our closed-coding process to adopt taxonomy categories may have introduced subjectivity. While grounded in prior work, different coding decisions could lead to slightly different categorization. We chose closed coding because prior studies~\cite{he2023python,wei2024demystifying} on API misuse provide taxonomies that can be readily adapted with minor modifications, making them suitable for our context. This approach allows us to capture patterns relevant to LLM-generated code without reinventing categories from scratch, though it carries a risk of overlap with existing classifications. Another potential threat concerns the breadth of our API completion tasks. We mitigated this by sampling 3,000 distinct API signatures each for Python and Java from The Stack, covering overloads and version-specific variants through static analysis. While this approach ensures diversity across frequently used APIs, the topical coverage of the underlying packages is difficult to determine, since open-source repositories often mix multiple domains and lack consistent metadata. As a result, our dataset may not fully capture the distribution of APIs across specific application areas, and we leave a more fine-grained analysis of domain coverage for future work.

\subsection{External Validity}
A significant threat to external validity is the generalizability of the findings to other programming languages, APIs, and LLMs. Our dataset is limited to Python and Java, focusing on specific API misuse categories. While these languages and categories are widely used and relevant, the results may not directly extend to other contexts, such as low-level languages or highly specialized APIs. Similarly, the LLMs evaluated for Dr.Fix in this study, including GPT-4o, Qwen2.5-Coder-Instruct, and Llama-3.1, represent a subset of available models. The effectiveness of Dr.Fix with other LLMs or on alternative programming datasets remains uncertain. Furthermore, our comparison with prior APR baselines (e.g., CodeBERT and CodeT5) is limited to Java and deep learning APIs, following the scope of that work. Hence, relative improvements may differ in other ecosystems. Moreover, because our dataset emphasizes frequently used APIs, the results may not fully generalize to rarely used or deprecated APIs, which could exhibit distinct misuse patterns or outdated usage behaviors. Finally, while our analysis is grounded in large-scale manual annotation, this process is extremely resource-consuming and limits the breadth of models and datasets we could feasibly include. As such, our findings may not generalize to other LLM families not studied here, and we leave broader cross-model validation to future work.
% Besides, there are some concerns regarding the temporal gap between the models studied for API completion (e.g., StarCoder-7B and Copilot) and the larger, more recent models (e.g., Llama-3.1 70B, Qwen2.5-Coder 32B, GPT-4o) evaluated with Dr.Fix. Our goal was to investigate whether today’s strongest LLMs can effectively detect and repair the typical API misuses produced by yesterday’s widely deployed IDE-scale models. This choice highlights a realistic adoption challenge: developers will continue to encounter code produced by earlier models even as newer, more capable models become available. While this design allows us to study repair in a forward-looking, cross-generation setting, it also means that the absolute error rates of newer models on the original API-completion task remain unknown, and future work should examine whether such models indeed commit fewer mistakes or different types of mistakes.

\subsection{Construct Validity}
\label{sec:construct_validity}
Construct validity may be threatened by the choice of evaluation metrics and dataset design. Metrics such as BLEU and CodeBERTScore, while widely used, primarily measure syntactic and semantic similarity and may not fully capture the functional correctness of the repairs. LLMs-as-Judges~\cite{zhuo2024ice, he2025code} is hence a more promising direction to explore for code quality evaluation. Moreover, the inclusion of 100 snippets without API misuse to evaluate refusal rates introduces a potential bias in the dataset composition. Although this addition aims to measure the models' ability to avoid hallucinating misuse, its proportion relative to actual misuse cases may influence the results. Finally, our use of human-written code from The Stack as the ground truth may embed biases or errors, and multiple valid API usages could be penalized under our evaluation setup. We also rely on the open-source community’s sustained effort to ensure the quality and relevance of The Stack, which may introduce variability beyond our control. In addition, the EM metric, while capturing structural matches at the AST level, may conflate syntactic similarity with semantic correctness, limiting its interpretability. Future work could explore alternative metrics or evaluation methods that better assess functional correctness and developer usability.

\section{Related Work}
\label{sec:related_work}

\subsection{Large Language Models for Code}
LLMs have emerged as a transformative approach for code generation tasks, broadly categorized into standard language models and instruction-tuned models. Standard language models are pre-trained on raw corpora using next-token prediction, exemplified by models such as Codex~\cite{chen2021evaluating}, CodeGen~\cite{nijkampcodegen}, CodeGeeX~\cite{zheng2023codegeex}, CodeT5~\cite{wang2021codet5} and StarCoder~\cite{li2023starcoder}. Codex, with 12B parameters, and StarCoder, featuring 15.5B parameters, demonstrated significant advancements in program synthesis through training on GitHub code and additional sources.
Instruction-tuned models represent the next evolution, with examples like ChatGPT employing Reinforcement Learning with Human Feedback~\cite{ouyang2022training}. Open-source alternatives include WizardCoder~\cite{luowizardcoder}, which fine-tunes StarCoder using techniques like Evol-Instruct, and InstructCodeT5+~\cite{wang2023codet5+}, which extends CodeT5 through instruction-tuned processes. These developments have enabled zero-shot capabilities in code generation tasks, demonstrating emergent behaviors that were previously unattainable while maintaining the essential citations and academic format.

\subsection{Bug Study of Large-Language-Model-Generated Code}
Understanding bugs in LLM-generated code is crucial for improving code generation capabilities. While prior studies have analyzed bugs in deep learning systems broadly~\cite{islam2019comprehensive,wang2022empirical,zhang2018empirical}, recent research has focused specifically on defects in LLM-generated code~\cite{liu2024refining}.
\cite{tambon2024bugs} investigated bugs in non-standalone code generated by LLMs, though their study was limited by potential data leakage issues and the use of now-outdated models. \cite{dou2024s} addressed these limitations by evaluating LLM-generated code with a robust benchmark designed to minimize data leakage, providing a more accurate representation of model performance in practical applications. Our work builds upon these insights by focusing specifically on API misuse patterns in LLM-generated code, introducing a taxonomy of error types and proposing the Dr.Fix repair mechanism.

\subsection{Program Repair by Large Language Models}
The success of LLMs in Natural Language Processing has led to their application in Automated Program Repair (APR)~\cite{jiang2021cure,xia2022less}. Key advancements include integrating LLMs with additional contextual information, such as TFix~\cite{berabi2021tfix}, which builds on the T5 model~\cite{raffel2020exploring} by incorporating error messages from diagnostic tools to improve repair accuracy.
Specialized models for code review and repair have also emerged~\cite{li2022automating,zhang2022coditt5}, along with zero-shot learning approaches that frame repair as a cloze-style task. Notable examples include Xia and Zhang's work~\cite{xia2022less} using masked language modeling and CIRCLE~\cite{yuan2022circle}'s prompt-based approach for code completion. Our work, Dr.Fix, builds upon these foundations by specifically addressing API misuse in generated code, combining detection, reasoning, and repair capabilities across Python and Java codebases.

\section{Conclusion and Future Work}
In this work, we systematically studied API misuse in LLM-generated code, analyzing thousands of code snippets across Python and Java. Through rigorous manual annotation, we developed a comprehensive taxonomy of API misuse types and quantified their prevalence in popular code generation models. Our findings reveal that LLMs struggle significantly with hallucination and intent misalignment, with distinct patterns emerging between programming languages. The shared error patterns between Copilot and StarCoder highlight fundamental challenges in API usage that persist across different model architectures.
We introduced Dr.Fix, a novel LLM-based APR approach that combines detection and reasoning capabilities to address these challenges. 
Our evaluation demonstrates that Dr.Fix significantly outperforms existing learning-based APR methods, substantially improving repair accuracy and maintaining high refusal rates for correct API usage. The results underscore the effectiveness of our staged approach in understanding and fixing API misuse.

Future work should focus on extending Dr.Fix's capabilities to handle more complex API misuse scenarios and investigating its applicability to other programming languages. Additionally, exploring the integration of Dr.Fix with development environments and investigating ways to improve its reasoning capabilities through enhanced prompting strategies could further advance automated API repair. We also plan to expand our taxonomy and dataset as new patterns of API misuse emerge with the evolution of code generation models.

\section*{Acknowledgements}

 TY.Zhuo is supported by CSIRO’s Data61 PhD Scholarships and IBM PhD Fellowship Awards. J.Grundy is supported by ARC Laureate Fellowship FL190100035. This research was supported by the Singapore Ministry of Education (MOE) Academic Research Fund (AcRF) Tier 1 grant (Project ID: 23-SOL-SMU-004). Any opinions, findings and conclusions or recommendations expressed in this material are those of the author(s) and do not reflect the views of the Ministry of Education, Singapore.

% \section{References}

\bibliographystyle{IEEEtran}
\bibliography{main}
\newpage
\begin{IEEEbiography}[{\includegraphics[width=1in,height=1.25in,clip]{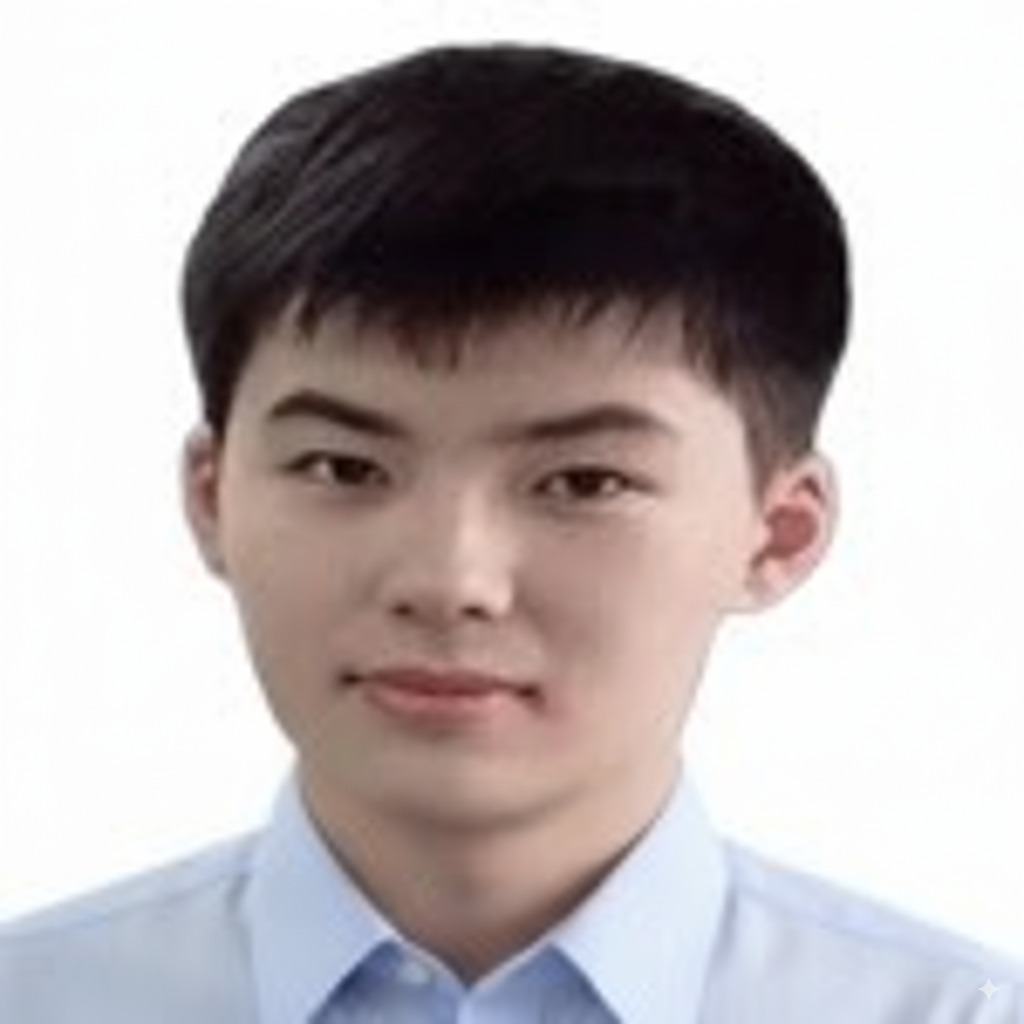}}]{Terry Yue Zhuo} a PhD student at Monash University and CSIRO's Data61, Australia. He previously worked at CSIRO's Data61, Singapore Management University, Sea AI Lab, TikTok AI Innovation Center, and Amazon AWS AI. He has authored more than 40 papers in refereed journals or conferences, such as The Web Conference (WWW), ACM Transactions on Software Engineering and Methodology (TOSEM), International Conference on Learning Representations (ICLR), Network and Distributed System Security (NDSS), Annual Meeting of the Association for Computational Linguistics (ACL) and Conference on Empirical Methods in Natural Language Processing (EMNLP). He received the best paper award in the Deep Learning for Code (DL4C) workshop at ICRL 2023, 2024-2025 IBM PhD Fellowship Awards, and 2025 DAAD AInet Fellowship. He serves as a Senior Area Chair of EMNLP 2025 and ACL 2026. His research interests include empirical software engineering, code intelligence, and security intelligence. More information about him can be found at \url{https://terryyz.github.io/}.
\end{IEEEbiography}

\begin{IEEEbiography}[{\includegraphics[width=1in,height=1.25in,clip]{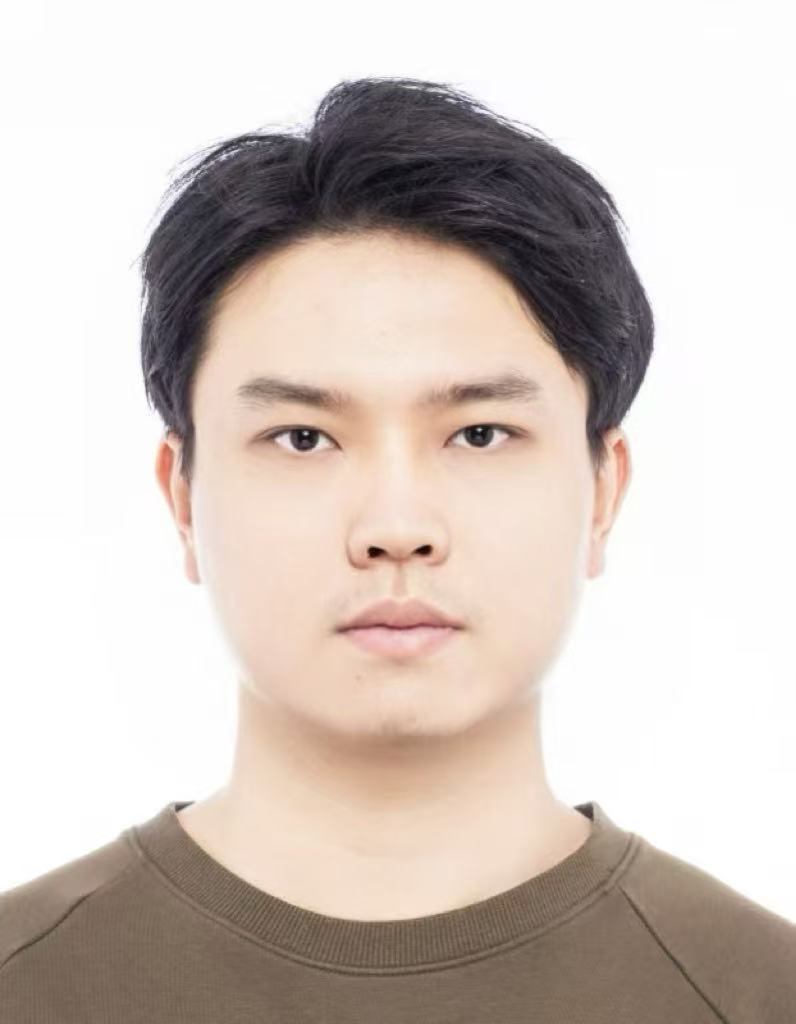}}]{Junda He} is a PhD candidate at the Singapore Management University, supervised by Professor David Lo. He earned his Master of Science degree in Software System Engineering and his Bachelor of Science degree in Computer Science, both from University College London. Junda's current research focuses on testing and enhancing the trustworthiness of AI agents. More information about him can be found at \url{https://jundahe.github.io/}.

\end{IEEEbiography}

\begin{IEEEbiography}
[{\includegraphics[width=1in,height=1.25in,clip]{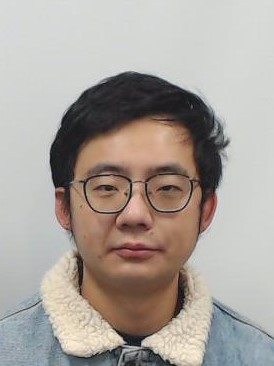}}] 
{Jiamou Sun} is a Research Scientist in the SE4AI team at CSIRO’s Data61, specialising in responsible AI, software supply chain security, and AI-driven automation. His research bridges artificial intelligence, cybersecurity, and software engineering, with a particular focus on knowledge graph–based vulnerability detection and risk assessment. He has led and contributed to numerous high-impact projects, including the CSIRO–Google Software Supply Chain Security initiative, the AISI Joint AI Safety Testing, VAISS co-editing, and Responsible AI for AI4M. His work has been recognised through publications in top-tier venues such as ICSE, ASE, WWW and UIST, as well as the distinguished paper awards in the ICSME'2018 and 2019. More information about him can be found at \url{https://scholar.google.com/citations?user=l2UCgDYAAAAJ}.
\end{IEEEbiography}

\begin{IEEEbiography}
[{\includegraphics[width=1in,height=1.25in,clip]{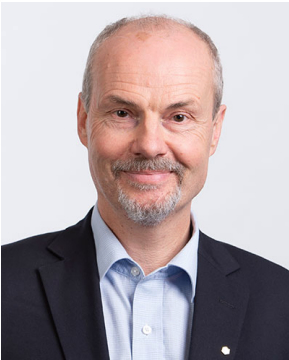}}] {John Grundy} is Australian Laureate Fellow, Senior Deputy Dean and
Professor of Software Engineering at Monash
University. He leads the HumaniSE lab in the
Faculty of Information Technology, investigating ``human-centric" issues in software engineering. These include impact of personality on software engineers and users; emotion-oriented requirements engineering; impact of different languages, cultures and belief sets on using and
engineering software; usability and accessibility
of software, particularly for ageing people and
people with physical and mental challenges; issues of gender, age,
socio-economic status and personal values on software, software requirements, and software engineering teams. He is IEEE Fellow, Fellow of Engineers Australia and Fellow of Automated Software Engineering. More information about him can be found at \url{https://sites.google.com/site/johncgrundy}.
\end{IEEEbiography}

\begin{IEEEbiography}
[{\includegraphics[width=1in,height=1.25in,clip]{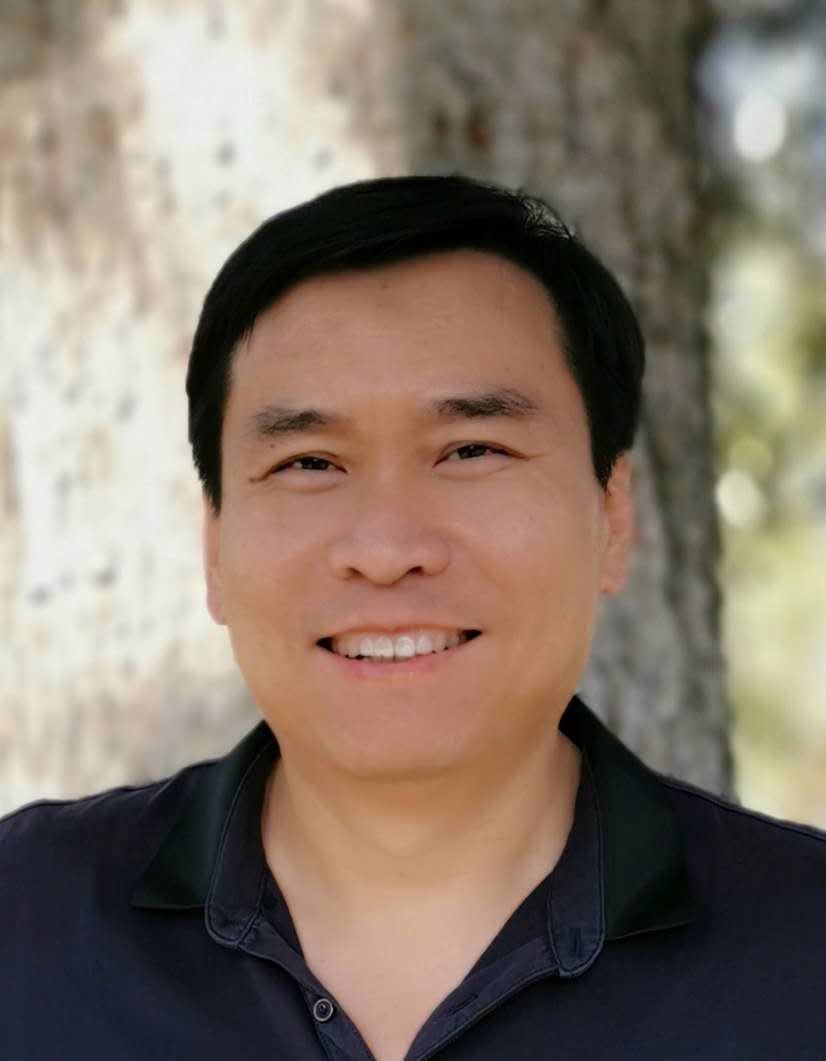}}] {Zhenchang Xing} is the Senior Principal Research Scientist at CSIRO's Data61. He is the Science Lead of Software System group. His research interests are Software Engineering, Responsible AI, and Human-Computer Interaction, with specific focus on the software engineering infrastructure for AI systems. His research received several ACM SIGSOFT Distinguished Paper Awards and IEEE TCSE Disginguished Paper Awards, as well as ACM SIGSOFT Most Influential Paper Award at ASE 2021. More information about him can be found at \url{https://people.csiro.au/X/Z/Zhenchang-Xing/}.
\end{IEEEbiography}

\begin{IEEEbiography}
[{\includegraphics[width=1in,height=1.25in,clip]{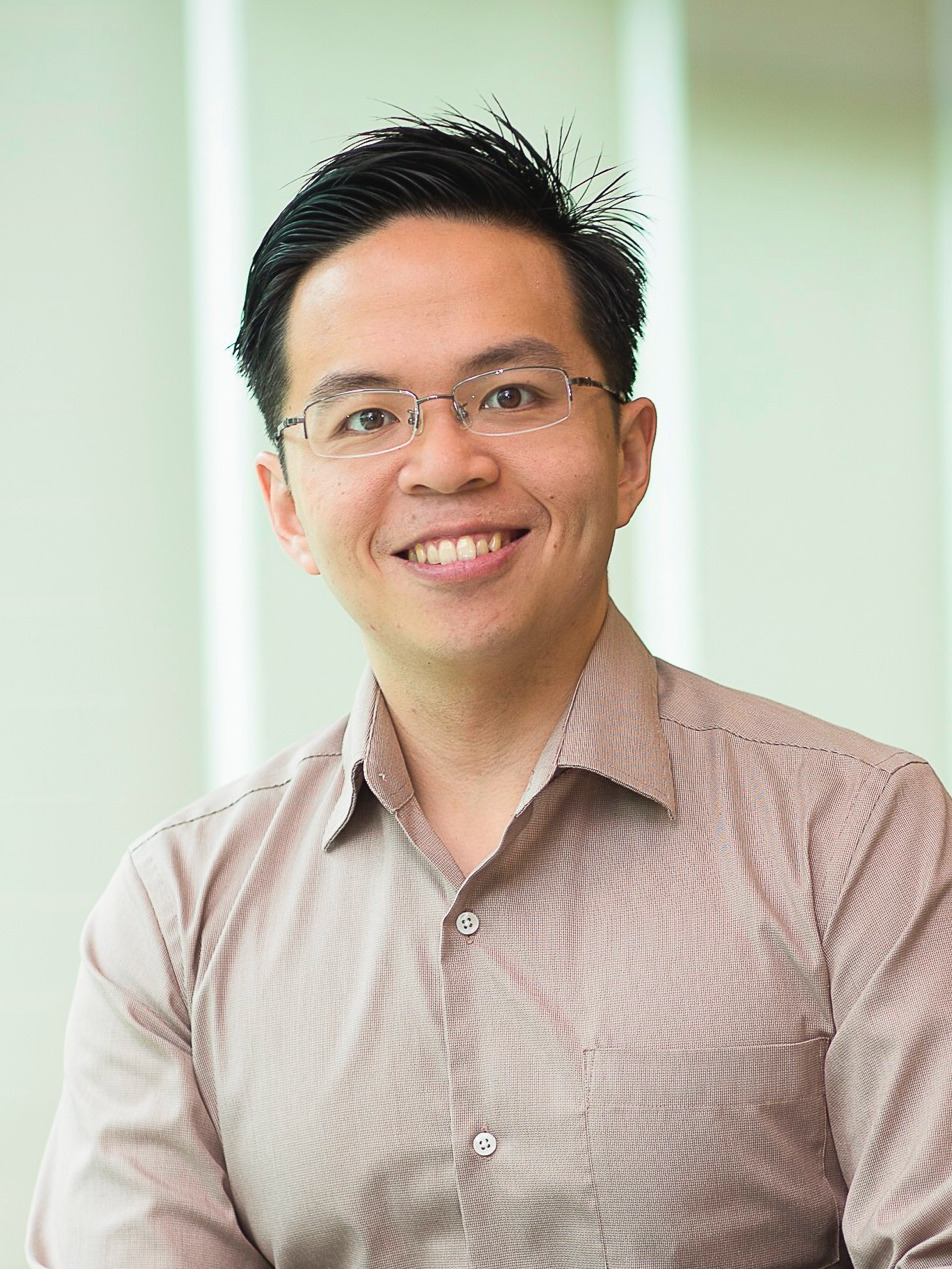}}] 
{David Lo} is the OUB Chair Professor of Computer Science and Founding Director of the Center for Research in Intelligent Software Engineering (RISE) at Singapore Management University. Championing the area of AI for Software Engineering (AI4SE) since the mid-2000s, he has demonstrated how AI — encompassing data mining, machine learning, information retrieval, natural language processing, and search-based algorithms — can transform software engineering data into actionable insights and automation. Through empirical studies, he has also identified practitioners' pain points, characterized the limitations of AI4SE solutions, and explored practitioners' acceptance thresholds for AI-powered tools. His contributions have led to over 20 awards, including four Test-of-Time awards and thirteen ACM SIGSOFT/IEEE TCSE Distinguished Paper awards, and his work has garnered over 40,000 citations. An ACM Fellow, IEEE Fellow, ASE Fellow, and National Research Foundation Investigator (Senior Fellow), Lo has also served as a PC Co-Chair for ASE'20, FSE'24, and ICSE'25. More information about him can be found at \url{http://www.mysmu.edu/faculty/davidlo/}.
\end{IEEEbiography}

\begin{IEEEbiography}
[{\includegraphics[width=1in,height=1.25in,clip]{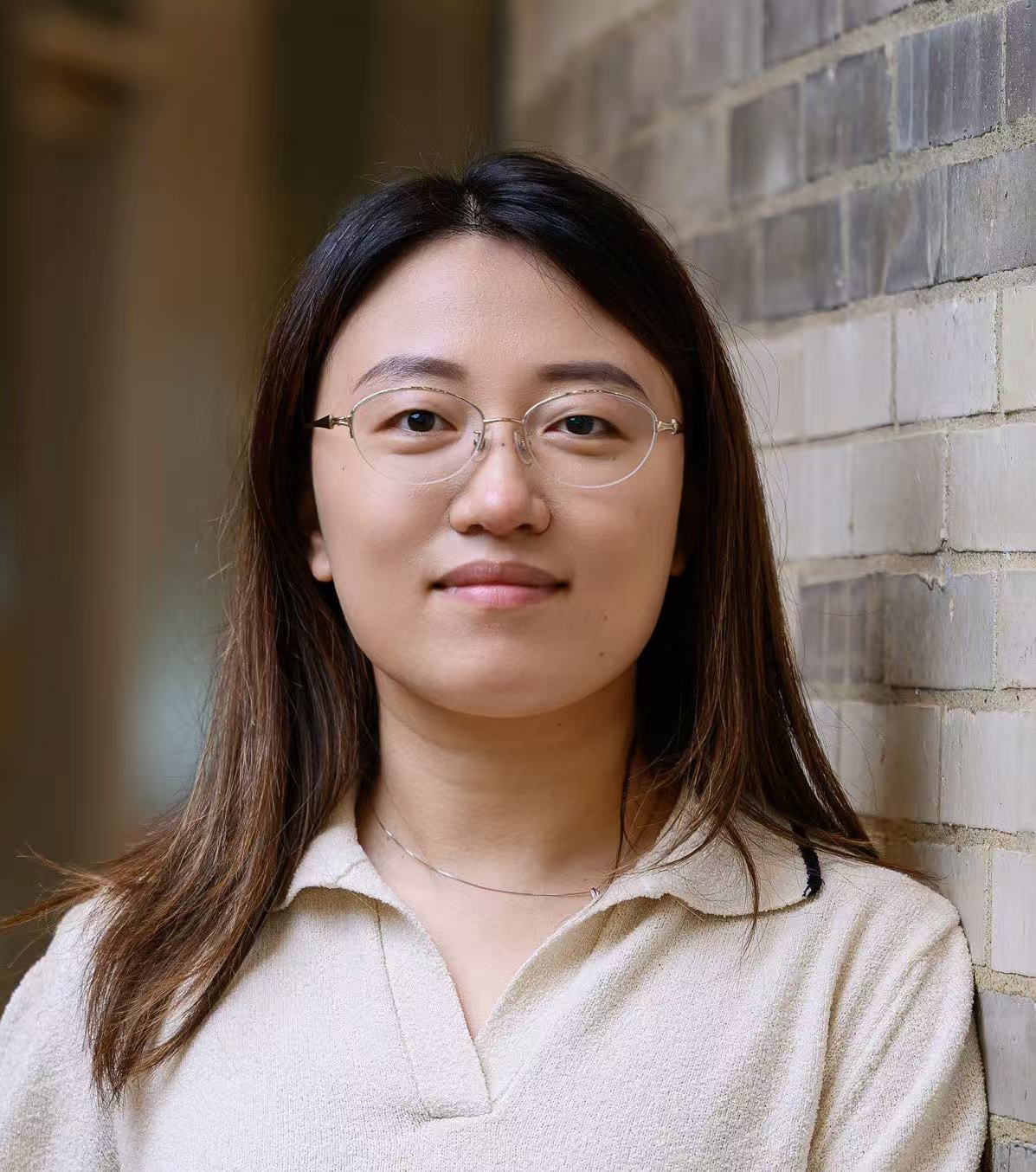}}] 
{Xiaoning Du} is an ARC DECRA Fellow and Senior Lecturer (equivalent to U.S. Associate Professor) at the Faculty of Information Technology, Monash University. She received her Ph.D. from Nanyang Technological University in 2020 and her Bachelor's degree from Fudan University in 2014. Her research primarily focuses on the security and quality assurance of intelligent software systems, with a particular emphasis on intelligent software engineering tools. She has published over 40 papers in top-tier conferences and journals. Her work has won or been nominated for the ACM SIGSOFT Distinguished Paper Award multiple times. She also received the 2024 FIT Dean's Early Career Researcher of the Year Award and the 2024 Google Research Scholar Award.
Check out more about her at \url{http:https://xiaoningdu.github.io/}.
\end{IEEEbiography}

% \newpage


\end{document}

%% file: def.tex
\usepackage[T1]{fontenc}    % use 8-bit T1 fonts
\usepackage{hyperref}       % hyperlinks
\usepackage{url}            % simple URL typesetting
\usepackage{booktabs}       % professional-quality tables
\usepackage{amsfonts}       % blackboard math symbols
\usepackage{nicefrac}       % compact symbols for 1/2, etc.
\usepackage{colortbl}         % colors
\usepackage{listings}
\usepackage{amsmath}
\usepackage{algorithm}
\usepackage{todonotes}
\usepackage{algpseudocode}
\usepackage{multirow}
\usepackage{arydshln}
\usepackage{tikz}
\usepackage{makecell}
\usepackage{pifont} 
\usepackage{placeins}
\usepackage{xspace}
\usepackage{mdframed}
\usepackage{ragged2e}

\definecolor{kellygreen}{rgb}{0.3, 0.73, 0.09}
\definecolor{alizarin}{rgb}{0.82, 0.1, 0.26}

\usepackage{amssymb} 
\colorlet{gray}{gray!20}
\colorlet{lightyellow}{yellow!20!white}
\colorlet{lightblue}{cyan!10!white}
\colorlet{lightgreen}{green!20!white}
\colorlet{lightred}{red!20!white}
\usepackage{listings}
\usepackage{mathtools}  % Add to preamble

\definecolor{codegreen}{rgb}{0,0.6,0}
\definecolor{codegray}{rgb}{0.5,0.5,0.5}
\definecolor{codepurple}{rgb}{0.58,0,0.82}
\definecolor{backcolour}{rgb}{0.95,0.95,0.92}

\definecolor{Gray}{gray}{0.9}
\definecolor{OliveGreen}{cmyk}{0.64,0,0.95,0.40}
\definecolor{lightlightgray}{gray}{0.9}
\usepackage[scaled=0.85]{beramono}
\definecolor{javared}{rgb}{0.6,0,0} % for strings
\definecolor{javagreen}{rgb}{0.25,0.5,0.35} % comments
\definecolor{javapurple}{rgb}{0.5,0,0.2} % keywords
\definecolor{javadocblue}{rgb}{0.25,0.35,0.75} % javadoc

\lstdefinelanguage{java-pretty}
{
  language=Python,
  numbers=left,
  frame=shadowbox,
  rulesepcolor= \color{red!20!green!20!blue!20},
  basicstyle=\footnotesize\ttfamily,
  numberstyle=\scriptsize,
  breaklines=true,
  columns=fullflexible,
  xleftmargin=16pt,
  showstringspaces=false,
  keywordstyle=\color{blue}\bfseries,
  stringstyle=\color{javared},
  commentstyle=\color{javagreen},
  morecomment=[s][\color{javadocblue}]{/**}{*/},
}

\lstdefinelanguage{diffpy}{
    language=Python,
    morecomment=[f][\color{diffstart}]{@@},
    morecomment=[f][\color{javagreen}]{+\ },
    morecomment=[f][\color{javared}]{-\ },
}

\lstdefinelanguage{diffjava}{
    language=Java,
    morecomment=[f][\color{diffstart}]{@@},
    morecomment=[f][\color{javagreen}]{+\ },
    morecomment=[f][\color{javared}]{-\ },
}

\definecolor{diffstart}{rgb}{0.4, 0.4, 0.4}

\newmdenv[
  backgroundcolor=gray!10,  % Light gray background
  linecolor=gray!80!black,   % Darker border color
  outerlinewidth=1pt,        % Border width
  innertopmargin=10pt,      % Top padding
  innerbottommargin=10pt,   % Bottom padding
  innerleftmargin=10pt,     % Left padding
  innerrightmargin=10pt,    % Right padding
  skipabove=10pt,           % Space above the box
  skipbelow=10pt            % Space below the box
]{findingbox}

%% file: main.bib
@inproceedings{zhong2024can,
  title={Can LLM Replace Stack Overflow? A Study on Robustness and Reliability of Large Language Model Code Generation},
  author={Zhong, Li and Wang, Zilong},
  booktitle={Proceedings of the AAAI Conference on Artificial Intelligence},
  volume={38},
  number={19},
  pages={21841--21849},
  year={2024}
}

@article{zhuo2024bigcodebench,
  title={Bigcodebench: Benchmarking code generation with diverse function calls and complex instructions},
  author={Zhuo, Terry Yue and Vu, Minh Chien and Chim, Jenny and Hu, Han and Yu, Wenhao and Widyasari, Ratnadira and Yusuf, Imam Nur Bani and Zhan, Haolan and He, Junda and Paul, Indraneil and others},
  journal={arXiv preprint arXiv:2406.15877},
  year={2024}
}

@article{touvron2023llama,
  title={Llama: Open and efficient foundation language models},
  author={Touvron, Hugo and Lavril, Thibaut and Izacard, Gautier and Martinet, Xavier and Lachaux, Marie-Anne and Lacroix, Timoth{\'e}e and Rozi{\`e}re, Baptiste and Goyal, Naman and Hambro, Eric and Azhar, Faisal and others},
  journal={arXiv preprint arXiv:2302.13971},
  year={2023}
}

@article{roziere2023code,
  title={Code llama: Open foundation models for code},
  author={Roziere, Baptiste and Gehring, Jonas and Gloeckle, Fabian and Sootla, Sten and Gat, Itai and Tan, Xiaoqing Ellen and Adi, Yossi and Liu, Jingyu and Sauvestre, Romain and Remez, Tal and others},
  journal={arXiv preprint arXiv:2308.12950},
  year={2023}
}

@article{chen2019sequencer,
  title={Sequencer: Sequence-to-sequence learning for end-to-end program repair},
  author={Chen, Zimin and Kommrusch, Steve and Tufano, Michele and Pouchet, Louis-No{\"e}l and Poshyvanyk, Denys and Monperrus, Martin},
  journal={IEEE Transactions on Software Engineering},
  volume={47},
  number={9},
  pages={1943--1959},
  year={2019},
  publisher={IEEE}
}

@misc{llm_api_misuse_replication,
  title        = {Replication Package},
  howpublished = {\url{https://github.com/terryyz/llm_api_misuse}},
  note         = {Replication package submitted for artifact evaluation},
  year         = {2025}
}

@inproceedings{zhu2021syntax,
  title={A syntax-guided edit decoder for neural program repair},
  author={Zhu, Qihao and Sun, Zeyu and Xiao, Yuan-an and Zhang, Wenjie and Yuan, Kang and Xiong, Yingfei and Zhang, Lu},
  booktitle={Proceedings of the 29th ACM joint meeting on European software engineering conference and symposium on the foundations of software engineering},
  pages={341--353},
  year={2021}
}

@article{achiam2023gpt,
  title={Gpt-4 technical report},
  author={Achiam, Josh and Adler, Steven and Agarwal, Sandhini and Ahmad, Lama and Akkaya, Ilge and Aleman, Florencia Leoni and Almeida, Diogo and Altenschmidt, Janko and Altman, Sam and Anadkat, Shyamal and others},
  journal={arXiv preprint arXiv:2303.08774},
  year={2023}
}

@article{li2023starcoder,
  title={StarCoder: May the Source be With You!},
  author={Li, R and Allal, LB and Zi, Y and Muennighoff, N and Kocetkov, D and Mou, C and Marone, M and Akiki, C and Li, J and Chim, J and others},
  journal={Transactions on machine learning research},
  year={2023},
  publisher={OpenReview}
}

@article{islah2024gitchameleon,
  title={GitChameleon: Unmasking the Version-Switching Capabilities of Code Generation Models},
  author={Islah, Nizar and Gehring, Justine and Misra, Diganta and Muller, Eilif and Rish, Irina and Zhuo, Terry Yue and Caccia, Massimo},
  journal={arXiv preprint arXiv:2411.05830},
  year={2024}
}

@article{kuhar2024libevolutioneval,
  title={LibEvolutionEval: A Benchmark and Study for Version-Specific Code Generation},
  author={Kuhar, Sachit and Ahmad, Wasi Uddin and Wang, Zijian and Jain, Nihal and Qian, Haifeng and Ray, Baishakhi and Ramanathan, Murali Krishna and Ma, Xiaofei and Deoras, Anoop},
  journal={arXiv preprint arXiv:2412.04478},
  year={2024}
}

@article{zhong2017empirical,
  title={An empirical study on API usages},
  author={Zhong, Hao and Mei, Hong},
  journal={IEEE Transactions on Software Engineering},
  volume={45},
  number={4},
  pages={319--334},
  year={2017},
  publisher={IEEE}
}

@inproceedings{mousavi2024investigation,
  title={An investigation into misuse of java security apis by large language models},
  author={Mousavi, Zahra and Islam, Chadni and Moore, Kristen and Abuadbba, Alsharif and Babar, M Ali},
  booktitle={Proceedings of the 19th ACM Asia Conference on Computer and Communications Security},
  pages={1299--1315},
  year={2024}
}

@article{nguyen2010graph,
  title={A graph-based approach to API usage adaptation},
  author={Nguyen, Hoan Anh and Nguyen, Tung Thanh and Wilson Jr, Gary and Nguyen, Anh Tuan and Kim, Miryung and Nguyen, Tien N},
  journal={ACM Sigplan Notices},
  volume={45},
  number={10},
  pages={302--321},
  year={2010},
  publisher={ACM New York, NY, USA}
}

@inproceedings{weng2023large,
  title={Large Language Models are Better Reasoners with Self-Verification},
  author={Weng, Yixuan and Zhu, Minjun and Xia, Fei and Li, Bin and He, Shizhu and Liu, Shengping and Sun, Bin and Liu, Kang and Zhao, Jun},
  booktitle={The 2023 Conference on Empirical Methods in Natural Language Processing}
}

@inproceedings{ren2020demystify,
  title={Demystify official API usage directives with crowdsourced API misuse scenarios, erroneous code examples and patches},
  author={Ren, Xiaoxue and Sun, Jiamou and Xing, Zhenchang and Xia, Xin and Sun, Jianling},
  booktitle={Proceedings of the ACM/IEEE 42nd international conference on software engineering},
  pages={925--936},
  year={2020}
}

@inproceedings{galappaththi2024empirical,
  title={An Empirical Study of API Misuses of Data-Centric Libraries},
  author={Galappaththi, Akalanka and Nadi, Sarah and Treude, Christoph},
  booktitle={Proceedings of the 18th ACM/IEEE International Symposium on Empirical Software Engineering and Measurement},
  pages={245--256},
  year={2024}
}

@inproceedings{wei2024demystifying,
  title={Demystifying and Detecting Misuses of Deep Learning APIs},
  author={Wei, Moshi and Harzevili, Nima Shiri and Huang, YueKai and Yang, Jinqiu and Wang, Junjie and Wang, Song},
  booktitle={Proceedings of the IEEE/ACM 46th International Conference on Software Engineering},
  pages={1--12},
  year={2024}
}

@article{kechagia2021evaluating,
  title={Evaluating automatic program repair capabilities to repair api misuses},
  author={Kechagia, Maria and Mechtaev, Sergey and Sarro, Federica and Harman, Mark},
  journal={IEEE Transactions on Software Engineering},
  volume={48},
  number={7},
  pages={2658--2679},
  year={2021},
  publisher={IEEE}
}

@article{zhang2023evaluating,
  title={Evaluating Pre-trained Language Models for Repairing API Misuses},
  author={Zhang, Ting and Irsan, Ivana Clairine and Thung, Ferdian and Lo, David and Sharma, Asankhaya and Jiang, Lingxiao},
  journal={arXiv preprint arXiv:2310.16390},
  year={2023}
}

@article{xu2022ide,
  title={In-ide code generation from natural language: Promise and challenges},
  author={Xu, Frank F and Vasilescu, Bogdan and Neubig, Graham},
  journal={ACM Transactions on Software Engineering and Methodology (TOSEM)},
  volume={31},
  number={2},
  pages={1--47},
  year={2022},
  publisher={ACM New York, NY}
}

@inproceedings{friedincoder,
  title={InCoder: A Generative Model for Code Infilling and Synthesis},
  author={Fried, Daniel and Aghajanyan, Armen and Lin, Jessy and Wang, Sida and Wallace, Eric and Shi, Freda and Zhong, Ruiqi and Yih, Scott and Zettlemoyer, Luke and Lewis, Mike},
  booktitle={The Eleventh International Conference on Learning Representations}
}

@inproceedings{kechagia2019effective,
  title={Effective and efficient API misuse detection via exception propagation and search-based testing},
  author={Kechagia, Maria and Devroey, Xavier and Panichella, Annibale and Gousios, Georgios and Van Deursen, Arie},
  booktitle={Proceedings of the 28th ACM SIGSOFT international symposium on software testing and analysis},
  pages={192--203},
  year={2019}
}

@article{kang2021active,
  title={Active learning of discriminative subgraph patterns for api misuse detection},
  author={Kang, Hong Jin and Lo, David},
  journal={IEEE Transactions on Software Engineering},
  volume={48},
  number={8},
  pages={2761--2783},
  year={2021},
  publisher={IEEE}
}

@inproceedings{li2021arbitrar,
  title={Arbitrar: User-guided api misuse detection},
  author={Li, Ziyang and Machiry, Aravind and Chen, Binghong and Naik, Mayur and Wang, Ke and Song, Le},
  booktitle={2021 IEEE Symposium on Security and Privacy (SP)},
  pages={1400--1415},
  year={2021},
  organization={IEEE}
}

@inproceedings{pradel2012statically,
  title={Statically checking API protocol conformance with mined multi-object specifications},
  author={Pradel, Michael and Jaspan, Ciera and Aldrich, Jonathan and Gross, Thomas R},
  booktitle={2012 34th International Conference on Software Engineering (ICSE)},
  pages={925--935},
  year={2012},
  organization={IEEE}
}

@inproceedings{pradel2012leveraging,
  title={Leveraging test generation and specification mining for automated bug detection without false positives},
  author={Pradel, Michael and Gross, Thomas R},
  booktitle={2012 34th International Conference on Software Engineering (ICSE)},
  pages={288--298},
  year={2012},
  organization={IEEE}
}

@article{hu2024active,
  title={Active Code Learning: Benchmarking Sample-Efficient Training of Code Models},
  author={Hu, Qiang and Guo, Yuejun and Xie, Xiaofei and Cordy, Maxime and Ma, Lei and Papadakis, Mike and Le Traon, Yves},
  journal={IEEE Transactions on Software Engineering},
  year={2024},
  publisher={IEEE}
}

@inproceedings{liu2023codegen4libs,
  title={Codegen4libs: A two-stage approach for library-oriented code generation},
  author={Liu, Mingwei and Yang, Tianyong and Lou, Yiling and Du, Xueying and Wang, Ying and Peng, Xin},
  booktitle={2023 38th IEEE/ACM International Conference on Automated Software Engineering (ASE)},
  pages={434--445},
  year={2023},
  organization={IEEE}
}

@inproceedings{ma2024compositional,
  title={Compositional API Recommendation for Library-Oriented Code Generation},
  author={Ma, Zexiong and An, Shengnan and Xie, Bing and Lin, Zeqi},
  booktitle={Proceedings of the 32nd IEEE/ACM International Conference on Program Comprehension},
  pages={87--98},
  year={2024}
}

@article{monperrus2013detecting,
  title={Detecting missing method calls as violations of the majority rule},
  author={Monperrus, Martin and Mezini, Mira},
  journal={ACM Transactions on Software Engineering and Methodology (TOSEM)},
  volume={22},
  number={1},
  pages={1--25},
  year={2013},
  publisher={ACM New York, NY, USA}
}

@article{radford2018improving,
  title={Improving language understanding by generative pre-training},
  author={Radford, Alec},
  year={2018}
}

@article{amann2018systematic,
  title={A systematic evaluation of static api-misuse detectors},
  author={Amann, Sven and Nguyen, Hoan Anh and Nadi, Sarah and Nguyen, Tien N and Mezini, Mira},
  journal={IEEE Transactions on Software Engineering},
  volume={45},
  number={12},
  pages={1170--1188},
  year={2018},
  publisher={IEEE}
}

@article{bush2000static,
  title={A static analyzer for finding dynamic programming errors},
  author={Bush, William R and Pincus, Jonathan D and Sielaff, David J},
  journal={Software: Practice and Experience},
  volume={30},
  number={7},
  pages={775--802},
  year={2000},
  publisher={Wiley Online Library}
}

@inproceedings{khatchadourian2020empirical,
  title={An Empirical Study on the Use and Misuse of Java 8 Streams.},
  author={Khatchadourian, Raffi and Tang, Yiming and Bagherzadeh, Mehdi and Ray, Baishakhi},
  booktitle={FASE},
  pages={97--118},
  year={2020}
}

@article{lamothe2021systematic,
  title={A systematic review of API evolution literature},
  author={Lamothe, Maxime and Gu{\'e}h{\'e}neuc, Yann-Ga{\"e}l and Shang, Weiyi},
  journal={ACM Computing Surveys (CSUR)},
  volume={54},
  number={8},
  pages={1--36},
  year={2021},
  publisher={ACM New York, NY}
}

@inproceedings{amann2016mubench,
  title={MUBench: A benchmark for API-misuse detectors},
  author={Amann, Sven and Nadi, Sarah and Nguyen, Hoan A and Nguyen, Tien N and Mezini, Mira},
  booktitle={Proceedings of the 13th international conference on mining software repositories},
  pages={464--467},
  year={2016}
}

@article{Wiggers2023StarCoderFreeModel,
  author = {Kyle Wiggers},
  title = {Hugging Face and ServiceNow Release a Free Code-Generating Model},
  journal = {TechCrunch},
  date = {2023-05-04},
  url = {https://techcrunch.com/2023/05/04/hugging-face-and-servicenow-release-a-free-code-generating-model/},
}

@article{IBM2023watsonxMainframe,
  author    = {IBM},
  title     = {IBM Unveils watsonx Generative AI Capabilities to Accelerate Mainframe Application Modernization},
  journal   = {IBM Newsroom},
  year      = {2023},
  month     = {August},
  day       = {22},
  url       = {}
}

@article{Dohmke2023CopilotTransformsPlatform,
  author    = {Thomas Dohmke},
  title     = {Universe 2023: Copilot Transforms GitHub into the AI-Powered Developer Platform},
  journal   = {GitHub Blog},
  year      = {2023},
  month     = {November},
  day       = {8},
  url       = {}
}

@inproceedings{sclarquantifying,
  title={Quantifying Language Models' Sensitivity to Spurious Features in Prompt Design or: How I learned to start worrying about prompt formatting},
  author={Sclar, Melanie and Choi, Yejin and Tsvetkov, Yulia and Suhr, Alane},
  booktitle={The Twelfth International Conference on Learning Representations}
}

@article{kwon2023exploring,
  title={Exploring LLM-based automated repairing of Ansible script in edge-cloud infrastructures},
  author={Kwon, Sunjae and Lee, Sungu and Kim, Taehyoun and Ryu, Duksan and Baik, Jongmoon},
  journal={Journal of web engineering},
  volume={22},
  number={6},
  pages={889--912},
  year={2023},
  publisher={River Publishers}
}

@inproceedings{izadi2024language,
  title={Language models for code completion: A practical evaluation},
  author={Izadi, Maliheh and Katzy, Jonathan and Van Dam, Tim and Otten, Marc and Popescu, Razvan Mihai and Van Deursen, Arie},
  booktitle={Proceedings of the IEEE/ACM 46th International Conference on Software Engineering},
  pages={1--13},
  year={2024}
}

@inproceedings{semenkin2025full,
  title={Full line code completion: Bringing ai to desktop},
  author={Semenkin, Anton and Bibaev, Vitaliy and Sokolov, Yaroslav and Krylov, Kirill and Kalina, Alexey and Khannanova, Anna and Savenkov, Danila and Rovdo, Darya and Davidenko, Igor and Karnaukhov, Kirill and others},
  booktitle={2025 IEEE/ACM 47th International Conference on Software Engineering: Software Engineering in Practice (ICSE-SEIP)},
  pages={563--574},
  year={2025},
  organization={IEEE}
}

@inproceedings{takerngsaksiri2024students,
  title={Students' perspectives on ai code completion: Benefits and challenges},
  author={Takerngsaksiri, Wannita and Warusavitarne, Cleshan and Yaacoub, Christian and Hou, Matthew Hee Keng and Tantithamthavorn, Chakkrit},
  booktitle={2024 IEEE 48th Annual Computers, Software, and Applications Conference (COMPSAC)},
  pages={1606--1611},
  year={2024},
  organization={IEEE}
}

@inproceedings{miaoselfcheck,
  title={SelfCheck: Using LLMs to Zero-Shot Check Their Own Step-by-Step Reasoning},
  author={Miao, Ning and Teh, Yee Whye and Rainforth, Tom},
  booktitle={The Twelfth International Conference on Learning Representations}
}

@article{mizrahi2024state,
  title={State of what art? a call for multi-prompt llm evaluation},
  author={Mizrahi, Moran and Kaplan, Guy and Malkin, Dan and Dror, Rotem and Shahaf, Dafna and Stanovsky, Gabriel},
  journal={Transactions of the Association for Computational Linguistics},
  volume={12},
  pages={933--949},
  year={2024},
  publisher={MIT Press 255 Main Street, 9th Floor, Cambridge, Massachusetts 02142, USA~…}
}

@inproceedings{raychev2014code,
  title={Code completion with statistical language models},
  author={Raychev, Veselin and Vechev, Martin and Yahav, Eran},
  booktitle={Proceedings of the 35th ACM SIGPLAN conference on programming language design and implementation},
  pages={419--428},
  year={2014}
}

@inproceedings{schlichtig2022fum,
  title={Fum-a framework for api usage constraint and misuse classification},
  author={Schlichtig, Michael and Sassalla, Steffen and Narasimhan, Krishna and Bodden, Eric},
  booktitle={2022 IEEE International Conference on Software Analysis, Evolution and Reengineering (SANER)},
  pages={673--684},
  year={2022},
  organization={IEEE}
}

@inproceedings{proksch2016evaluating,
  title={Evaluating the evaluations of code recommender systems: a reality check},
  author={Proksch, Sebastian and Amann, Sven and Nadi, Sarah and Mezini, Mira},
  booktitle={Proceedings of the 31st IEEE/ACM International Conference on Automated Software Engineering},
  pages={111--121},
  year={2016}
}

@online{openrouter,
  author       = {{OpenRouter}},
  title        = {OpenRouter: Unified Access to Large Language Models},
  year         = {2024},
  url          = {https://openrouter.ai/},
  note         = {Accessed: 2025-10-13}
}

@inproceedings{he2023python,
  title={How Dynamic Features Affect API Usages? An Empirical Study of API Misuses in Python Programs},
  author={He, Xincheng and Liu, Xiaojin and Xu, Lei and Xu, Baowen},
  booktitle={2023 IEEE International Conference on Software Analysis, Evolution and Reengineering (SANER)},
  pages={522--533},
  year={2023},
  organization={IEEE}
}

@inproceedings{zhauo-etal-2025-nlp,
    title = "{NLP}+{C}ode: Code Intelligence in Language Models",
    author = "Zhuo, Terry Yue  and
      Liu, Qian  and
      Wang, Zijian  and
      Ahmad, Wasi Uddin  and
      Hui, Binyuan  and
      Allal, Loubna Ben",
    editor = "Pyatkin, Valentina  and
      Vlachos, Andreas",
    booktitle = "Proceedings of the 2025 Conference on Empirical Methods in Natural Language Processing: Tutorial Abstracts",
    month = nov,
    year = "2025",
    address = "Suzhou, China",
    publisher = "Association for Computational Linguistics",
    url = "https://aclanthology.org/2025.emnlp-tutorials.4/",
    doi = "10.18653/v1/2025.emnlp-tutorials.4",
    pages = "9--11",
    ISBN = "979-8-89176-336-4"
}

@inproceedings{
muennighoff2024octopack,
title={OctoPack: Instruction Tuning Code Large Language Models},
author={Niklas Muennighoff and Qian Liu and Armel Randy Zebaze and Qinkai Zheng and Binyuan Hui and Terry Yue Zhuo and Swayam Singh and Xiangru Tang and Leandro Von Werra and Shayne Longpre},
booktitle={The Twelfth International Conference on Learning Representations},
year={2024},
url={https://openreview.net/forum?id=mw1PWNSWZP}
}

@inproceedings{svyatkovskiy2020intellicode,
  title={Intellicode compose: Code generation using transformer},
  author={Svyatkovskiy, Alexey and Deng, Shao Kun and Fu, Shengyu and Sundaresan, Neel},
  booktitle={Proceedings of the 28th ACM joint meeting on European software engineering conference and symposium on the foundations of software engineering},
  pages={1433--1443},
  year={2020}
}

@inproceedings{fu2024serverlessllm,
  title={ServerlessLLM: Low-latency serverless inference for large language models},
  author={Fu, Yao and Xue, Leyang and Huang, Yeqi and Brabete, Andrei-Octavian and Ustiugov, Dmitrii and Patel, Yuvraj and Mai, Luo},
  booktitle={18th USENIX Symposium on Operating Systems Design and Implementation},
  pages={135--153},
  year={2024},
  organization={USENIX Association}
}

@article{kocetkovstack,
  title={The Stack: 3 TB of permissively licensed source code},
  author={Kocetkov, Denis and Li, Raymond and Jia, LI and Mou, Chenghao and Jernite, Yacine and Mitchell, Margaret and Ferrandis, Carlos Mu{\~n}oz and Hughes, Sean and Wolf, Thomas and Bahdanau, Dzmitry and others},
  journal={Transactions on Machine Learning Research}
}

@inproceedings{papineni2002bleu,
  title={Bleu: a method for automatic evaluation of machine translation},
  author={Papineni, Kishore and Roukos, Salim and Ward, Todd and Zhu, Wei-Jing},
  booktitle={Proceedings of the 40th annual meeting of the Association for Computational Linguistics},
  pages={311--318},
  year={2002}
}

@inproceedings{zhou2023codebertscore,
  title={CodeBERTScore: Evaluating Code Generation with Pretrained Models of Code},
  author={Zhou, Shuyan and Alon, Uri and Agarwal, Sumit and Neubig, Graham},
  booktitle={Proceedings of the 2023 Conference on Empirical Methods in Natural Language Processing},
  pages={13921--13937},
  year={2023}
}

@inproceedings{zhangbertscore,
  title={BERTScore: Evaluating Text Generation with BERT},
  author={Zhang, Tianyi and Kishore, Varsha and Wu, Felix and Weinberger, Kilian Q and Artzi, Yoav},
  booktitle={International Conference on Learning Representations}
}

@article{mchugh2012interrater,
  title={Interrater reliability: the kappa statistic},
  author={McHugh, Mary L},
  journal={Biochemia medica},
  volume={22},
  number={3},
  pages={276--282},
  year={2012},
  publisher={Medicinska naklada}
}

@article{dubey2024llama,
  title={The llama 3 herd of models},
  author={Dubey, Abhimanyu and Jauhri, Abhinav and Pandey, Abhinav and Kadian, Abhishek and Al-Dahle, Ahmad and Letman, Aiesha and Mathur, Akhil and Schelten, Alan and Yang, Amy and Fan, Angela and others},
  journal={arXiv preprint arXiv:2407.21783},
  year={2024}
}

@article{hui2024qwen2,
  title={Qwen2. 5-coder technical report},
  author={Hui, Binyuan and Yang, Jian and Cui, Zeyu and Yang, Jiaxi and Liu, Dayiheng and Zhang, Lei and Liu, Tianyu and Zhang, Jiajun and Yu, Bowen and Lu, Keming and others},
  journal={arXiv preprint arXiv:2409.12186},
  year={2024}
}

@article{cassano2023type,
  title={Type Prediction With Program Decomposition and Fill-in-the-Type Training},
  author={Cassano, Federico and Yee, Ming-Ho and Shinn, Noah and Guha, Arjun and Holtzen, Steven},
  journal={arXiv preprint arXiv:2305.17145},
  year={2023}
}

@inproceedings{peng2023generative,
  title={Generative type inference for python},
  author={Peng, Yun and Wang, Chaozheng and Wang, Wenxuan and Gao, Cuiyun and Lyu, Michael R},
  booktitle={2023 38th IEEE/ACM International Conference on Automated Software Engineering (ASE)},
  pages={988--999},
  year={2023},
  organization={IEEE}
}

@article{chen2021evaluating,
  title={Evaluating large language models trained on code},
  author={Chen, Mark and Tworek, Jerry and Jun, Heewoo and Yuan, Qiming and Pinto, Henrique Ponde De Oliveira and Kaplan, Jared and Edwards, Harri and Burda, Yuri and Joseph, Nicholas and Brockman, Greg and others},
  journal={arXiv preprint arXiv:2107.03374},
  year={2021}
}

@inproceedings{nijkampcodegen,
  title={CodeGen: An Open Large Language Model for Code with Multi-Turn Program Synthesis},
  author={Nijkamp, Erik and Pang, Bo and Hayashi, Hiroaki and Tu, Lifu and Wang, Huan and Zhou, Yingbo and Savarese, Silvio and Xiong, Caiming},
  booktitle={The Eleventh International Conference on Learning Representations}
}

@inproceedings{zheng2023codegeex,
  title={Codegeex: A pre-trained model for code generation with multilingual benchmarking on humaneval-x},
  author={Zheng, Qinkai and Xia, Xiao and Zou, Xu and Dong, Yuxiao and Wang, Shan and Xue, Yufei and Shen, Lei and Wang, Zihan and Wang, Andi and Li, Yang and others},
  booktitle={Proceedings of the 29th ACM SIGKDD Conference on Knowledge Discovery and Data Mining},
  pages={5673--5684},
  year={2023}
}

@article{ouyang2022training,
  title={Training language models to follow instructions with human feedback},
  author={Ouyang, Long and Wu, Jeffrey and Jiang, Xu and Almeida, Diogo and Wainwright, Carroll and Mishkin, Pamela and Zhang, Chong and Agarwal, Sandhini and Slama, Katarina and Ray, Alex and others},
  journal={Advances in neural information processing systems},
  volume={35},
  pages={27730--27744},
  year={2022}
}

@inproceedings{luowizardcoder,
  title={WizardCoder: Empowering Code Large Language Models with Evol-Instruct},
  author={Luo, Ziyang and Xu, Can and Zhao, Pu and Sun, Qingfeng and Geng, Xiubo and Hu, Wenxiang and Tao, Chongyang and Ma, Jing and Lin, Qingwei and Jiang, Daxin},
  booktitle={The Twelfth International Conference on Learning Representations}
}

@inproceedings{wang2023codet5+,
  title={CodeT5+: Open Code Large Language Models for Code Understanding and Generation},
  author={Wang, Yue and Le, Hung and Gotmare, Akhilesh and Bui, Nghi and Li, Junnan and Hoi, Steven},
  booktitle={Proceedings of the 2023 Conference on Empirical Methods in Natural Language Processing},
  pages={1069--1088},
  year={2023}
}

@inproceedings{wang2021codet5,
  title={CodeT5: Identifier-aware Unified Pre-trained Encoder-Decoder Models for Code Understanding and Generation},
  author={Wang, Yue and Wang, Weishi and Joty, Shafiq and Hoi, Steven CH},
  booktitle={Proceedings of the 2021 Conference on Empirical Methods in Natural Language Processing},
  year={2021},
  organization={Association for Computational Linguistics}
}

@article{he2025code,
  title={From Code to Courtroom: LLMs as the New Software Judges},
  author={He, Junda and Shi, Jieke and Zhuo, Terry Yue and Treude, Christoph and Sun, Jiamou and Xing, Zhenchang and Du, Xiaoning and Lo, David},
  journal={arXiv preprint arXiv:2503.02246},
  year={2025}
}

@inproceedings{zhuo2024ice,
  title={ICE-Score: Instructing Large Language Models to Evaluate Code},
  author={Zhuo, Terry Yue},
  booktitle={Findings of the Association for Computational Linguistics: EACL 2024},
  pages={2232--2242},
  year={2024}
}

@inproceedings{islam2019comprehensive,
  title={A comprehensive study on deep learning bug characteristics},
  author={Islam, Md Johirul and Nguyen, Giang and Pan, Rangeet and Rajan, Hridesh},
  booktitle={Proceedings of the 2019 27th ACM joint meeting on european software engineering conference and symposium on the foundations of software engineering},
  pages={510--520},
  year={2019}
}

@inproceedings{wang2022empirical,
  title={An empirical study on numerical bugs in deep learning programs},
  author={Wang, Gan and Wang, Zan and Chen, Junjie and Chen, Xiang and Yan, Ming},
  booktitle={Proceedings of the 37th IEEE/ACM International Conference on Automated Software Engineering},
  pages={1--5},
  year={2022}
}

@inproceedings{zhang2018empirical,
  title={An empirical study on tensorflow program bugs},
  author={Zhang, Yuhao and Chen, Yifan and Cheung, Shing-Chi and Xiong, Yingfei and Zhang, Lu},
  booktitle={Proceedings of the 27th ACM SIGSOFT international symposium on software testing and analysis},
  pages={129--140},
  year={2018}
}

@article{tambon2024bugs,
  title={Bugs in large language models generated code},
  author={Tambon, Florian and Dakhel, Arghavan Moradi and Nikanjam, Amin and Khomh, Foutse and Desmarais, Michel C and Antoniol, Giuliano},
  journal={arXiv preprint arXiv:2403.08937},
  year={2024}
}

@article{dou2024s,
  title={What's Wrong with Your Code Generated by Large Language Models? An Extensive Study},
  author={Dou, Shihan and Jia, Haoxiang and Wu, Shenxi and Zheng, Huiyuan and Zhou, Weikang and Wu, Muling and Chai, Mingxu and Fan, Jessica and Huang, Caishuang and Tao, Yunbo and others},
  journal={arXiv preprint arXiv:2407.06153},
  year={2024}
}

@article{lozhkov2024starcoder,
  title={Starcoder 2 and the stack v2: The next generation},
  author={Lozhkov, Anton and Li, Raymond and Allal, Loubna Ben and Cassano, Federico and Lamy-Poirier, Joel and Tazi, Nouamane and Tang, Ao and Pykhtar, Dmytro and Liu, Jiawei and Wei, Yuxiang and others},
  journal={arXiv preprint arXiv:2402.19173},
  year={2024}
}

@inproceedings{sainz2023nlp,
  title={NLP Evaluation in trouble: On the Need to Measure LLM Data Contamination for each Benchmark},
  author={Sainz, Oscar and Campos, Jon Ander and Garc{\'\i}a-Ferrero, Iker and Etxaniz, Julen and de Lacalle, Oier Lopez and Agirre, Eneko},
  booktitle={The 2023 Conference on Empirical Methods in Natural Language Processing}
}

@article{liu2024refining,
  title={Refining chatgpt-generated code: Characterizing and mitigating code quality issues},
  author={Liu, Yue and Le-Cong, Thanh and Widyasari, Ratnadira and Tantithamthavorn, Chakkrit and Li, Li and Le, Xuan-Bach D and Lo, David},
  journal={ACM Transactions on Software Engineering and Methodology},
  volume={33},
  number={5},
  pages={1--26},
  year={2024},
  publisher={ACM New York, NY}
}

@inproceedings{jiang2021cure,
  title={Cure: Code-aware neural machine translation for automatic program repair},
  author={Jiang, Nan and Lutellier, Thibaud and Tan, Lin},
  booktitle={2021 IEEE/ACM 43rd International Conference on Software Engineering (ICSE)},
  pages={1161--1173},
  year={2021},
  organization={IEEE}
}

@inproceedings{berabi2021tfix,
  title={Tfix: Learning to fix coding errors with a text-to-text transformer},
  author={Berabi, Berkay and He, Jingxuan and Raychev, Veselin and Vechev, Martin},
  booktitle={International Conference on Machine Learning},
  pages={780--791},
  year={2021},
  organization={PMLR}
}

@article{brown2020language,
  title={Language models are few-shot learners},
  author={Brown, Tom and Mann, Benjamin and Ryder, Nick and Subbiah, Melanie and Kaplan, Jared D and Dhariwal, Prafulla and Neelakantan, Arvind and Shyam, Pranav and Sastry, Girish and Askell, Amanda and others},
  journal={Advances in neural information processing systems},
  volume={33},
  pages={1877--1901},
  year={2020}
}

@inproceedings{li2022automating,
  title={Automating code review activities by large-scale pre-training},
  author={Li, Zhiyu and Lu, Shuai and Guo, Daya and Duan, Nan and Jannu, Shailesh and Jenks, Grant and Majumder, Deep and Green, Jared and Svyatkovskiy, Alexey and Fu, Shengyu and others},
  booktitle={Proceedings of the 30th ACM Joint European Software Engineering Conference and Symposium on the Foundations of Software Engineering},
  pages={1035--1047},
  year={2022}
}

@inproceedings{zhang2022coditt5,
  title={Coditt5: Pretraining for source code and natural language editing},
  author={Zhang, Jiyang and Panthaplackel, Sheena and Nie, Pengyu and Li, Junyi Jessy and Gligoric, Milos},
  booktitle={Proceedings of the 37th IEEE/ACM International Conference on Automated Software Engineering},
  pages={1--12},
  year={2022}
}

@inproceedings{xia2022less,
  title={Less training, more repairing please: revisiting automated program repair via zero-shot learning},
  author={Xia, Chunqiu Steven and Zhang, Lingming},
  booktitle={Proceedings of the 30th ACM Joint European Software Engineering Conference and Symposium on the Foundations of Software Engineering},
  pages={959--971},
  year={2022}
}

@inproceedings{yuan2022circle,
  title={CIRCLE: continual repair across programming languages},
  author={Yuan, Wei and Zhang, Quanjun and He, Tieke and Fang, Chunrong and Hung, Nguyen Quoc Viet and Hao, Xiaodong and Yin, Hongzhi},
  booktitle={Proceedings of the 31st ACM SIGSOFT international symposium on software testing and analysis},
  pages={678--690},
  year={2022}
}

@inproceedings{
xu2024rejection,
title={Rejection Improves Reliability: Training {LLM}s to Refuse Unknown Questions Using {RL} from Knowledge Feedback},
author={Hongshen Xu and Zichen Zhu and Situo Zhang and Da Ma and Shuai Fan and Lu Chen and Kai Yu},
booktitle={First Conference on Language Modeling},
year={2024},
url={https://openreview.net/forum?id=lJMioZBoR8}
}

@article{zhang2023survey,
  title={A survey on learning to reject},
  author={Zhang, Xu-Yao and Xie, Guo-Sen and Li, Xiuli and Mei, Tao and Liu, Cheng-Lin},
  journal={Proceedings of the IEEE},
  volume={111},
  number={2},
  pages={185--215},
  year={2023},
  publisher={IEEE}
}

@inproceedings{zhang2024r,
  title={R-Tuning: Instructing Large Language Models to Say ‘I Don’t Know’},
  author={Zhang, Hanning and Diao, Shizhe and Lin, Yong and Fung, Yi and Lian, Qing and Wang, Xingyao and Chen, Yangyi and Ji, Heng and Zhang, Tong},
  booktitle={Proceedings of the 2024 Conference of the North American Chapter of the Association for Computational Linguistics: Human Language Technologies (Volume 1: Long Papers)},
  pages={7106--7132},
  year={2024}
}

@article{liu2024exploring,
  title={Exploring and evaluating hallucinations in llm-powered code generation},
  author={Liu, Fang and Liu, Yang and Shi, Lin and Huang, Houkun and Wang, Ruifeng and Yang, Zhen and Zhang, Li and Li, Zhongqi and Ma, Yuchi},
  journal={arXiv preprint arXiv:2404.00971},
  year={2024}
}

@article{zhang2024llm,
  title={Llm hallucinations in practical code generation: Phenomena, mechanism, and mitigation},
  author={Zhang, Ziyao and Wang, Yanlin and Wang, Chong and Chen, Jiachi and Zheng, Zibin},
  journal={arXiv preprint arXiv:2409.20550},
  year={2024}
}

@inproceedings{chenteaching,
  title={Teaching Large Language Models to Self-Debug},
  author={Chen, Xinyun and Lin, Maxwell and Sch{\"a}rli, Nathanael and Zhou, Denny},
  booktitle={The Twelfth International Conference on Learning Representations}
}

@article{raffel2020exploring,
  title={Exploring the limits of transfer learning with a unified text-to-text transformer},
  author={Raffel, Colin and Shazeer, Noam and Roberts, Adam and Lee, Katherine and Narang, Sharan and Matena, Michael and Zhou, Yanqi and Li, Wei and Liu, Peter J},
  journal={Journal of machine learning research},
  volume={21},
  number={140},
  pages={1--67},
  year={2020}
}

@article{tian2024codehalu,
  title={CodeHalu: Code Hallucinations in LLMs Driven by Execution-based Verification},
  author={Tian, Yuchen and Yan, Weixiang and Yang, Qian and Chen, Qian and Wang, Wen and Luo, Ziyang and Ma, Lei},
  journal={arXiv preprint arXiv:2405.00253},
  year={2024}
}
